\documentclass[10pt,conference,anonymous]{IEEEtran}
\IEEEoverridecommandlockouts
\usepackage[utf8]{inputenc}
\usepackage{array}
\usepackage{svg}
\usepackage{tabularx}
\usepackage{enumitem}
\usepackage[linesnumbered,ruled,vlined]{algorithm2e}
\usepackage{cite}
\usepackage{amsmath,amssymb,amsfonts}
\usepackage{wrapfig}
\usepackage{algorithmic}
\usepackage{graphicx}
\usepackage{tcolorbox}
\usepackage{booktabs}
\usepackage{textcomp}
\DeclareUnicodeCharacter{2212}{\textminus}
\usepackage{textcomp}
\usepackage{pifont}
\usepackage{subcaption}
\usepackage[table]{xcolor}
\usepackage{booktabs}
\usepackage{multirow}
\usepackage{url}
\usepackage[hidelinks]{hyperref}
\usepackage{tikz}
\usetikzlibrary{positioning, arrows.meta, fit, calc, shapes.geometric}
\usepackage{xspace}

\graphicspath{{Figures/}{Tables/}}
\newcommand{\mcg}{MCG\xspace}
\newcommand{\fcg}{FCG\xspace}
\newcommand{\chg}{CHG\xspace}
\newcommand{\pdg}{PDG\xspace}

\newcommand{\kai}[1]{\textcolor{black}{{#1}}}
\newcommand{\jy}[1]{\textcolor{black}{{#1}}}

\newif\ifshowcomments
\showcommentsfalse

\ifshowcomments
  \newcommand{\chen}[1]{\textcolor{orange}{#1}}
\else
  \newcommand{\chen}[1]{}
\fi

\newcommand{\TOOL}{\textsc{DualView}\xspace} %TOOL name

\newcommand{\TOOLat}[1]{\textsc{DualView}\textsubscript{@#1}}

\newcommand{\TOOLfull}{\TOOL{}$_{\textit{full}}$}
\newcommand{\TOOLbase}{\TOOL{}$_{\textit{base}}$}
\newcommand{\TOOLtextual}{\TOOL{}$_{\textit{textual}}$}
\newcommand{\TOOLvisual}{\TOOL{}$_{\textit{visual}}$}

\newcommand{\TOOLw}[1]{\TOOL{}$_{\textit{w/\,#1}}$}

\newcommand{\llm}{LLM\xspace}

\definecolor{myLightBlue}{RGB}{242,242,253}    % 背景色（很浅）
\definecolor{myLightGreen}{RGB}{242,251,242}   % very light green
\definecolor{myLightYellow}{RGB}{255,250,235}  % very light yellow
\definecolor{myLightOrange}{RGB}{255,244,235}  % very light orange
\definecolor{myLightPurple}{RGB}{248,242,255}  % very light purple
\definecolor{myLightGray}{RGB}{245,246,248}    % very light gray
\definecolor{myLightPink}{RGB}{255,242,248}    % very light pink
\definecolor{myLightBlueGray}{RGB}{244,244,251}

\newcommand{\kimiktwofive}{Kimi K2.5\xspace}

\newcommand{\claudesonnetfourfive}{Claude 4.5 Sonnet\xspace}

\newcommand{\anthropic}{\includegraphics[height=\fontcharht\font`C]{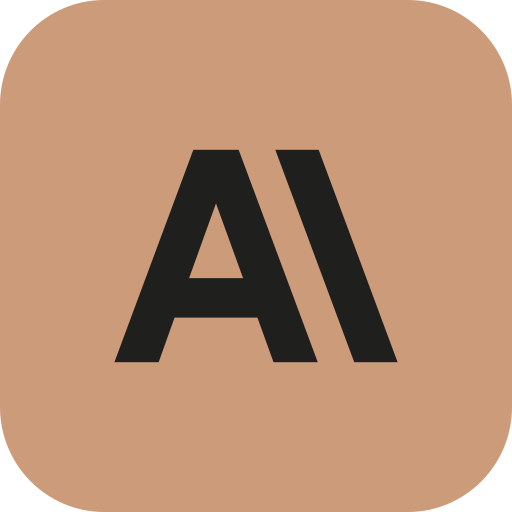}\xspace}

\newcommand{\kimi}{\includegraphics[height=\fontcharht\font`C]{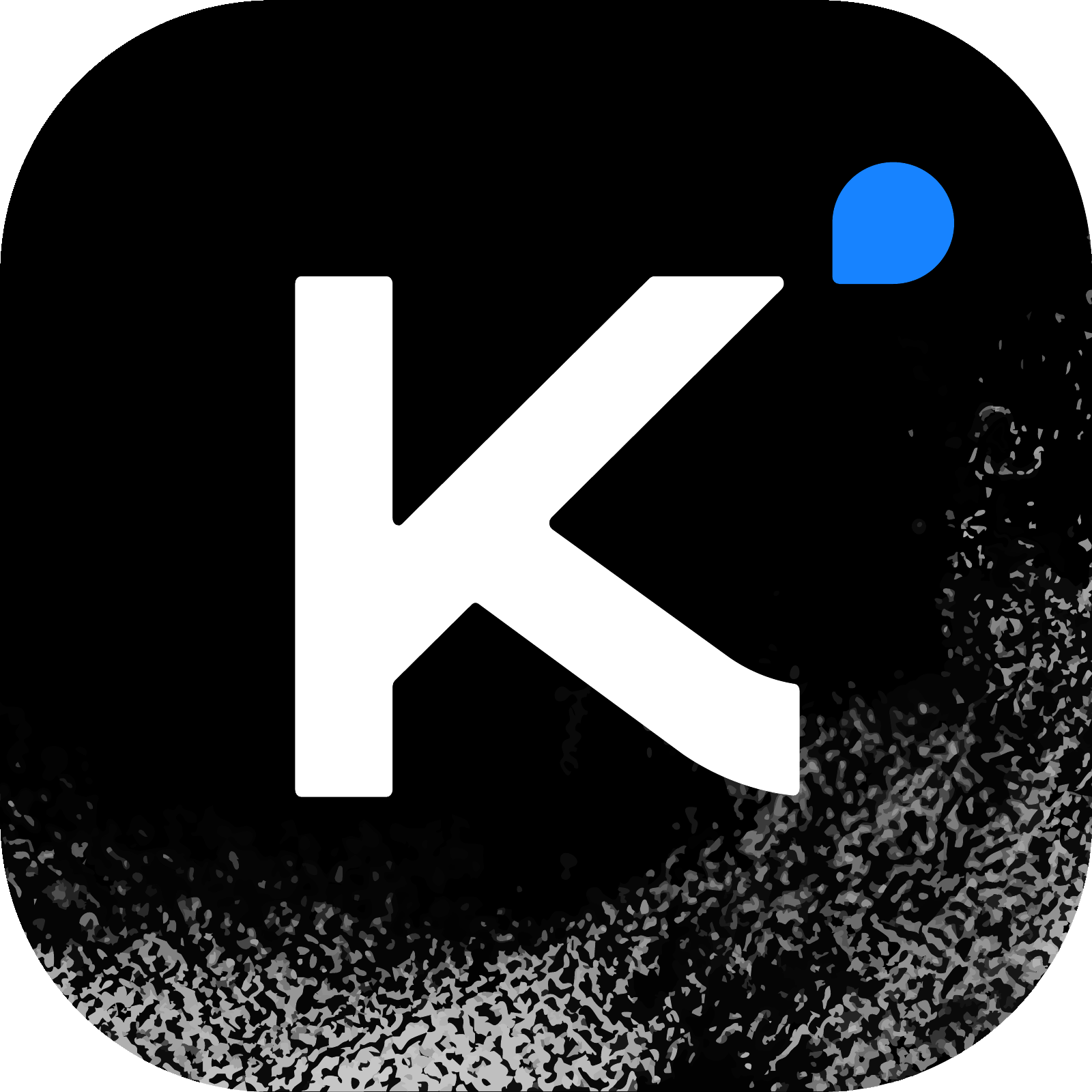}\xspace}

\newcommand{\swebv}{SWE-bench Verified\xspace}
\newcommand{\swebp}{SWE-bench Pro\xspace}

\newcommand{\swea}{SWE-agent\xspace}
\newcommand{\miniswea}{mini-SWE-agent\xspace}
\newcommand{\liveswea}{Live-SWE-agent\xspace}

\newcommand{\gain}[1]{{\tiny\textcolor[RGB]{111, 66, 193}{$\uparrow$#1}}}
\newcommand{\save}[1]{{\tiny\textcolor[RGB]{34, 139, 34}{$\downarrow$#1}}}

\definecolor{box0color}{RGB}{255, 0, 0}    % 红色 - 
\definecolor{box1color}{RGB}{230, 74, 25}    %  橙色 
\definecolor{box2color}{RGB}{255, 193, 7}    % 黄色 
\definecolor{box3color}{RGB}{13, 110, 253}   % 蓝色 
\definecolor{box4color}{RGB}{111, 66, 193}   % 紫色 
\definecolor{box7color}{RGB}{25, 135, 84}    % 绿色 
\definecolor{box8color}{RGB}{255, 193, 7}    % 黄色 

\definecolor{myLightBlue}{RGB}{242,242,253}   % 背景色（很浅）
\definecolor{myBlueLine}{RGB}{128,127,255}    % 左竖线颜色

\newtcolorbox{answerbox_round}{
  colback=myLightBlue,
  colframe=myBlueLine,
  boxrule=0pt,
  leftrule=3pt,
  arc=2pt,
  left=0pt,
  right=0pt,
  top=1pt,
  bottom=1pt,
  before skip=4pt,           % 盒子前的间距
  after skip=4pt,            % 盒子后的间距
}

\def\BibTeX{{\rm B\kern-.05em{\sc i\kern-.025em b}\kern-.08em
    T\kern-.1667em\lower.7ex\hbox{E}\kern-.125emX}}
\begin{document}

% 我认为目前最匹配你论文的故事
% 其实是：

% Limitation 1
% 文本代码探索难以感知结构
% ↓
% Limitation 2
% 文本图关系难以感知长程依赖
% ↓
% Solution
% Dual-modal graph representations
% Visual modality → topology / navigation
% Textual modality → details / semantics
% ↓
% Result
% Best performance comes from combining both

% \title{Seeing is Understanding: Visual Code-Graph Scaffolding for Issue-Resolution Agents}
% \title{Seeing is Understanding: Visual Structural Reasoning for Agentic Program Repair}

% \kai{ICSE Title:}
% \title{Seeing is Understanding: Dual-Modal Structural Reasoning for Agentic Program Repair}
% \kai{arXiv Title:}
\title{Beyond Textual Repository Exploration: Dual-Modal Structural Reasoning for Agentic Issue Resolution}

\author{\IEEEauthorblockN{Jiayi Zhang}
\IEEEauthorblockA{\textit{Nanyang Technological University} \\
% \textit{Nanyang Technological University}\\
% Singapore \\
jiayi043@e.ntu.edu.sg}
\and
\IEEEauthorblockN{Kai Huang\IEEEauthorrefmark{1}\thanks{\IEEEauthorrefmark{1}Corresponding author: Kai Huang (kai-kevin.huang@tum.de).}
}
\IEEEauthorblockA{\textit{TU Munich} \\
kai-kevin.huang@tum.de}
\and
\IEEEauthorblockN{Yang Liu}
\IEEEauthorblockA{\textit{Nanyang Technological University} \\
yangliu@ntu.edu.sg}
\and
\IEEEauthorblockN{Chunyang Chen}
\IEEEauthorblockA{\textit{TU Munich} \\
chun-yang.chen@tum.de}
}

% \author{Anonymous Author(s)}

\maketitle

\begin{abstract}

% Recent advances in agentic software engineering have significantly improved automated program repair by enabling iterative repository exploration. 
Recent advances in agentic program repair have significantly improved issue resolution by enabling iterative repository exploration. 
% Recent advances in agentic program repair have shown promising results in addressing repository-level software issues.
However, existing approaches predominantly rely on sequential, text-based code navigation, which fundamentally limits their ability to reason over large-scale long-horizon repositories with complex and long-range dependencies. As issue-resolution agents traverse repositories through fragmented textual observations, structural information such as module organization, call relationships, and dependency chains must be repeatedly reconstructed across interaction steps, often leading to exploration drift and incomplete localization.
We present \TOOL, a dual-modal structural scaffolding framework that brings visual reasoning into repository exploration for issue-resolution agents. \TOOL represents repository structure through four complementary graph views: Module Coupling Graph (MCG), Function Call Graph (FCG), Class Hierarchy Graph (CHG), and Program Dependence Graph (PDG), and exposes them through a queryable interface with visual and textual responses. Rather than reconstructing repository structure from a sequence of textual observations, agents can directly reason over persistent visual representations of code dependencies, enabling more effective exploration and understanding of long-horizon codebases.
We evaluate \TOOL on SWE-bench Pro and Verified.
% across multiple state-of-the-art issue-resolution agents and frontier LLMs. 
Results show that \TOOL consistently improves issue-resolution performance across different agent architectures and model families. Further ablation studies demonstrate that the gains arise not only from textual structural information but also from visual externalization of repository dependencies, which better supports long-horizon repository exploration.

\end{abstract}

\begin{IEEEkeywords}
Program repair, issue resolution, multimodal reasoning, code graphs, autonomous agents
\end{IEEEkeywords}

\section{Introduction}
\label{sec:intro}

\kai{Recent advances in large language model (LLM)-based agents have substantially advanced automated program repair (APR), enabling autonomous resolution of software issues in real-world repositories~\cite{SWE_Agent,OpenHands,yao2023reactsynergizingreasoningacting,zhang2024autocoderover,RepairAgent,ReinFix,Live_Swe_Agent,trae2025traeagent,Agentless}. Given an issue report, these agents iteratively explore the repository, identify the root cause, and synthesize a patch, achieving increasingly strong performance on repository-level issue resolution~\cite{SWEBench,SWEBench_Pro,SWEBench_M}. }

\begin{figure}[t]
    \centering
    % \vspace{-13pt}
    % \includegraphics[width=\linewidth]{Figures/motivation.pdf}
    % \includegraphics[width=\linewidth, trim=20 36 20 55, clip]{Figures/motivation_text_vs_dualmodal.pdf}
    \includegraphics[width=\linewidth, trim=21 34 21 49, clip]{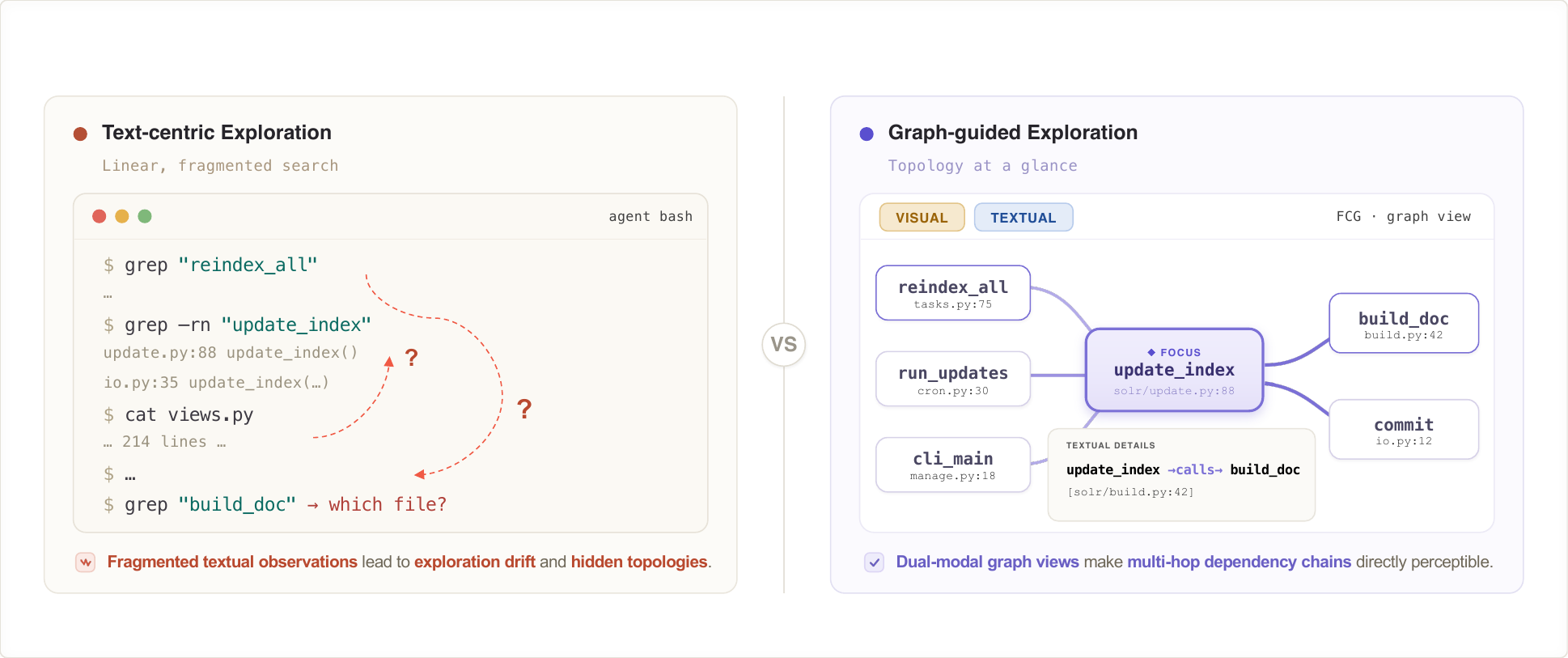}
    \caption{\textcolor[HTML]{B44F35}{Text-centric} vs. \textcolor{box4color}{Graph-guided} repository exploration. \chen{Both Figure 1 and 2 are too small to see clearly. Try to make the font size larger.} 
    % \kai{have updated figure}
    }
    \label{fig:motivation}
    \vspace{-14pt}
\end{figure}

\kai{Existing agent systems for issue resolution typically rely on the textual interface for repository exploration. For instance, on SWE-Bench Verified~\cite{SWEBenchVerfied}, simply reading repository files (via tools such as \textit{grep} and \textit{cat}) accounts for 76.1\% of a coding agent's token budget~\cite{SWE_Pruner}, indicating that text-based repository exploration dominates the agent's workload.}\chen{Is there any statistic e.g., for coding agent, how much time/token/tool calls are used for going through the repo? Tell the importance of that.} In practice, some refine and extend richer file-system and navigation primitives (e.g., purpose-built file viewers, search, and edit commands) to make code legible to the agent~\cite{SWE_Agent,OpenHands}.
Recent work such as mini-SWE-agent lets the model drive repository exploration through raw shell commands~\cite{mini_SWE_agent}.
Another line replaces free-form exploration with structured retrieval over program structure (e.g., abstract syntax trees, hierarchical localization, or repository-level code graphs) whose results are serialized back to the model as text~\cite{zhang2024autocoderover,Agentless,RepoGraph,CodeGraph,Prometheus}. 
Despite these advances, they share a common underlying assumption: that the agent can reconstruct and maintain an accurate understanding of repository structure purely from a sequence of textual observations.
\kai{As illustrated in Figure~\ref{fig:motivation}, existing agents explore repositories through a sequence of textual tool invocations, gradually piecing together structural relationships from fragmented observations. Although repository dependencies are inherently graph-structured, they are typically presented as linear text, leaving the underlying topology implicit. Representing these dependencies directly as graph-structured observations instead makes connectivity and multi-hop relationships immediately perceptible while preserving the semantic information required for subsequent source-level inspection.}
\chen{Want one figure in the first (prefered) or second page to show the difference of original text based unstructured code files (left-hand side), and nice visualization of graph (one kind on the right-hand side) to illustrate the intuition behind this work. Of course, the task should be single enough, and graph cover most relevant info. May move some pictures in the experiments to here. But try to get one nice figure for reader's first glance.}

\kai{Therefore}, 
the text-only interface imposes fundamental limitations on issue resolution.
% \kai: I revised two limitations based your suggestions
% \textcolor{box4color}{\textbf{Limitation} {\textbf{\ding{182}}}}: \textit{Repository relationships are difficult to perceive through textual \kai{code} inspection.}
% Repository-level issues often require reasoning about code dependencies that span modules, functions, classes, and statements. Although textual tools can expose individual files, symbols, or relations, they present this information sequentially. When the relevant structure contains multiple branches, cross-file paths, or densely connected components, the agent must reconstruct the underlying topology from separate observations. As a result, relationships such as call propagation, inheritance paths, and module coupling remain less directly observable than the source snippets in which they are encoded.
\textcolor{box4color}{\textbf{Limitation} {\textbf{\ding{182}}}}:
\textit{Repository exploration relies on fragmented textual observations.}
\kai{Issue-resolution agents typically explore repositories through textual tools such as file inspection, keyword search, and shell commands~\cite{SWE_Agent,zhang2024autocoderover,mini_SWE_agent}. Each interaction reveals only a small fragment of the repository, requiring the agent to gradually accumulate scattered observations across many tool invocations. Since repository relationships span modules, functions, classes, and statements, agents must mentally reconstruct these structural connections from sequential textual evidence, making repository exploration increasingly difficult as the search space grows.}
% \kai
% \textcolor{box4color}{\textbf{Limitation} {\textbf{\ding{183}}}}: \textit{Long-range dependency chains are difficult to reason through textual \kai{graph} representations.}
% Issue resolution often requires reasoning over dependency paths that span multiple functions, classes, modules, or control-flow regions. 
% While such relationships can be extracted by existing analysis tools and presented to the agent, they are typically serialized as textual descriptions or lists of edges. Although these textual representations preserve fine-grained semantic details, they linearize graph-structured dependencies into long sequences of tokens. As dependency structures become larger and more interconnected, critical paths and intermediate nodes become increasingly difficult to identify within lengthy textual outputs. 
% Consequently, agents may overlook important links in a dependency chain or fail to recognize higher-level structural patterns, making long-range reasoning over repository relationships both costly and error-prone.
\textcolor{box4color}{\textbf{Limitation} {\textbf{\ding{183}}}}:
\textit{Graph-structured repository information is still presented as text.}
\kai{Recent work has begun incorporating structural analyses such as repository graphs, call graphs, and hierarchical localization to guide repository exploration~\cite{RepoGraph,CodeGraph,ARISE}. However, these structural relationships are almost exclusively serialized into textual descriptions or edge lists before being passed to the agent. Although such representations preserve semantic information, they linearize graph topology into sequences of tokens. As dependency structures become deeper and more interconnected, multi-hop paths, branching structures, and densely connected regions become increasingly difficult to perceive, limiting the agent's ability to reason over long-range repository dependencies.}
% Together, these limitations arise from \textit{a mismatch between how repository structure is naturally organized and how it is presented to issue-resolution agents}. Repository relationships are inherently graph-structured, yet they are predominantly exposed through sequential textual observations that emphasize local details while obscuring global topology.
\kai{Together, these limitations stem from a common design choice: existing 
% issue-resolution agents 
agent systems
interact with graph-structured repositories 
% almost exclusively 
primarily
through textual interfaces. Whether exploring repositories via sequential tool invocations or consuming graph analyses serialized as text, agents must reconstruct repository topology from textual observations before reasoning over it. This mismatch between graph-structured repository and text-centric interaction makes long-horizon repository exploration both inefficient and error-prone.}

Recent advances in multimodal LLMs (MLLMs) demonstrate strong visual reasoning capabilities~\cite{DeepSeek_OCR,Kimi_Visual_Agentic_Intelligence,Agentic_Vision_in_Gemini}. Repository structure is naturally graph-structured, with dependencies among modules, functions, classes, and statements, yet existing agents primarily observe these relationships as sequential text. Visualizing repository graphs makes connectivity, hierarchies, and multi-hop dependency paths directly perceptible, enabling structural reasoning that is difficult to achieve from fragmented textual observations alone. This motivates a shift from text-centric repository exploration toward visual structural reasoning for issue resolution.

% -----------------------------

% To this end, we propose \TOOL, a multi-grained dual-modal scaffolding framework that equips agent systems with complementary visual views of repository structure. 
% \TOOL addresses the above challenge through two key design principles. First, it externalizes repair-relevant repository relationships into visual graph representations, making structural dependencies directly observable rather than requiring agents to reconstruct them from fragmented textual observations. Second, it preserves complex dependency structures in their native graph form, enabling agents to reason over long-range dependency chains and higher-level structural patterns that are often difficult to perceive from textual representations alone.

To this end, we propose \TOOL, a dual-modal scaffolding framework that equips 
agent systems
with multi-grained structural views of a repository.
\TOOL addresses the above challenges through two key design principles.
First, it externalizes repair-relevant repository relationships as graph-based structural abstractions, making dependencies directly accessible rather than requiring agents to reconstruct them from fragmented textual observations.
Second, it exposes these structures through complementary visual and textual representations. The visual modality preserves topology and long-range structural patterns, while the textual modality provides precise semantic details, enabling agents to jointly reason about repository organization and dependency propagation.

\TOOL decomposes the repository structure along 
% two orthogonal axes, granularity (coarse versus fine) and abstraction (organization versus interaction), yielding 
four classical software-engineering (SE) graphs: Module Coupling Graph (\mcg), Function Call Graph (\fcg), Class Hierarchy Graph (\chg), and Program Dependence Graph (\pdg). 
% The \mcg is rendered once at session start to provide a persistent global map of repository organization, while the remaining 
These graph views are exposed as parameterized structural probes that agents can invoke on demand. 
% Each query returns a dual-modal response consisting of a rendered graph image and a concise textual summary, enabling targeted structural exploration throughout the repair process. 
Each query returns a dual-modal response consisting of a rendered visual image and a concise textual summary, allowing agents to combine visual inspection of structural patterns with textual inspection of semantic details during repository exploration.

To evaluate the effectiveness of \TOOL, we integrate it into multiple representative agent systems and evaluate on both \swebp~\cite{SWEBench_Pro} and SWE-bench Verified~\cite{SWEBenchVerfied}. Across different agent architectures and frontier MLLMs, \TOOL consistently improves issue-resolution performance, demonstrating that dual-modal structural reasoning serves as a general enhancement rather than a model-specific optimization. On \swebp, \TOOL achieves up to 388 resolved instances, improving OpenCode with Kimi K2.5 by 46 additional resolved instances. Ablation studies further show that the gains arise not only from exposing repository structure, but also from how that structure is represented: visualized graphs outperform equivalent textual graph descriptions, while combining visual and textual graph representations yields the strongest performance. These findings suggest that dual-modal structural reasoning provides an effective interface for long-horizon repository exploration and dependency analysis.

% Specifically, we make the following contributions.

% \begin{itemize}[leftmargin=0.3cm]
% \item \textbf{Paradigm.}
% % 我们为program repair领域引入了新的修复范式，即突破传统的文本化代码库探索理解形式，转向迈向多模态的视觉化代码库依赖关系建模理解。
% We introduce a visual-interactive paradigm for repository exploration in program repair, where structural dependencies are externalized into visual artifacts that guide agentic reasoning.

% \item \textbf{Modality.} We introduce rendered repository structure as a first-class visual input channel for issue-resolution agents, distinct from prior multimodal software-engineering work that consumes user-supplied images.

% \item \textbf{Framework.} We propose a decomposition of repository representations along granularity and abstraction axes, mapping four classical software-engineering graphs (\mcg, \fcg, \chg, \pdg) onto a single coordinate system that mirrors the diagnostic progression of fault localization.

% \item \textbf{Evaluation.} On \swebp with Gemini 3.1 Pro, \TOOL resolves \TBD{N} issues, \TBD{+M} more than the strongest \miniswea baseline; an ablation isolates the marginal contribution of each graph and of the visual modality itself, and a generalizability study reproduces the gain on \swebv and on alternative \llms.

% \end{itemize}

% \kai:我们需要至少在主实验中跑两个模型，例如Gemini 3 Pro 和 Kimi 2.6, 因为它们都强调visual reasoning能力，并且官方的report直接提供了它们在SWE-Bench Pro上的结果，可以作为我们的直接基线，我们只要比它们高就行了

Specifically, we make the following contributions.

\begin{itemize}[leftmargin=*]

    \item \textbf{Dual-modal structural scaffolding framework.}
    We propose \TOOL, a dual-modal structural scaffolding framework that externalizes repository structure into synchronized visual and textual graph representations. By integrating four complementary structural views, \TOOL enables agents to reason across both repository-level organization and fine-grained program dependencies.
    % We propose \TOOL, a dual-modal multi-view framework that externalizes repository structure into visual and textual graph representations. By organizing four complementary SE graphs across multiple granularities, \TOOL enables agents to navigate and reason about both global-level repository organization and fine-grained program dependencies.

    \item \textbf{Comprehensive evaluation.}
    We evaluate \TOOL across multiple agent frameworks, foundation models, and benchmarks. Results demonstrate consistent improvements in issue-resolution performance, while ablation studies confirm the complementary benefits of multi-grained structural views and dual-modal structural observations.
    % We conduct comprehensive experiments across multiple agent frameworks, foundation models, and benchmarks. The results demonstrate that \TOOL consistently improves issue-resolution performance, while ablation studies validate the complementary contributions of the proposed multi-grained structural views and dual-modal structural observations.

    % \item \textbf{Open Access infrastructure.\chen{access}}
    \item \textbf{Open-source implementation.} 
    We release the implementation of \TOOL
    and further package the dual-modal structural reasoning interface as reusable tools and MCP services, to facilitate future research on multimodal repository reasoning and issue-resolution agents~\cite{Dualview_link,Dualview_MCP_repo}. \chen{Any detailed usage statistics?}
    
    % \item \textbf{Agentic visual reasoning interface.}
    % We design a queryable visual interaction mechanism in which issue-resolution agents dynamically invoke structural probes and receive dual-modal responses consisting of rendered graph visualizations and textual summaries, enabling targeted structural exploration during long-horizon repair trajectories.

    % \item \textbf{Open infrastructure.}
    % We release the implementation of \TOOL
    % and further package the dual-modal structural reasoning interface as reusable tools and MCP services, to facilitate future research on multimodal repository reasoning and issue-resolution agents.

    % 集成到opencode看看开源的反馈。MCP + PR

    % \kai{我们开源了Visual structural scaffolding framework \tool的所有代码实现，并且将Agentic visual reasoning interface封装成MCP工具提供给开源社区使用以促进未来的issue resolution agent社区的进一步发展未来的研究人员}

\end{itemize}
\section{Approach}
\label{sec:approach}

\begin{figure*}[t]
    \centering
    \includegraphics[width=\linewidth, trim=32 95 32 65, clip]{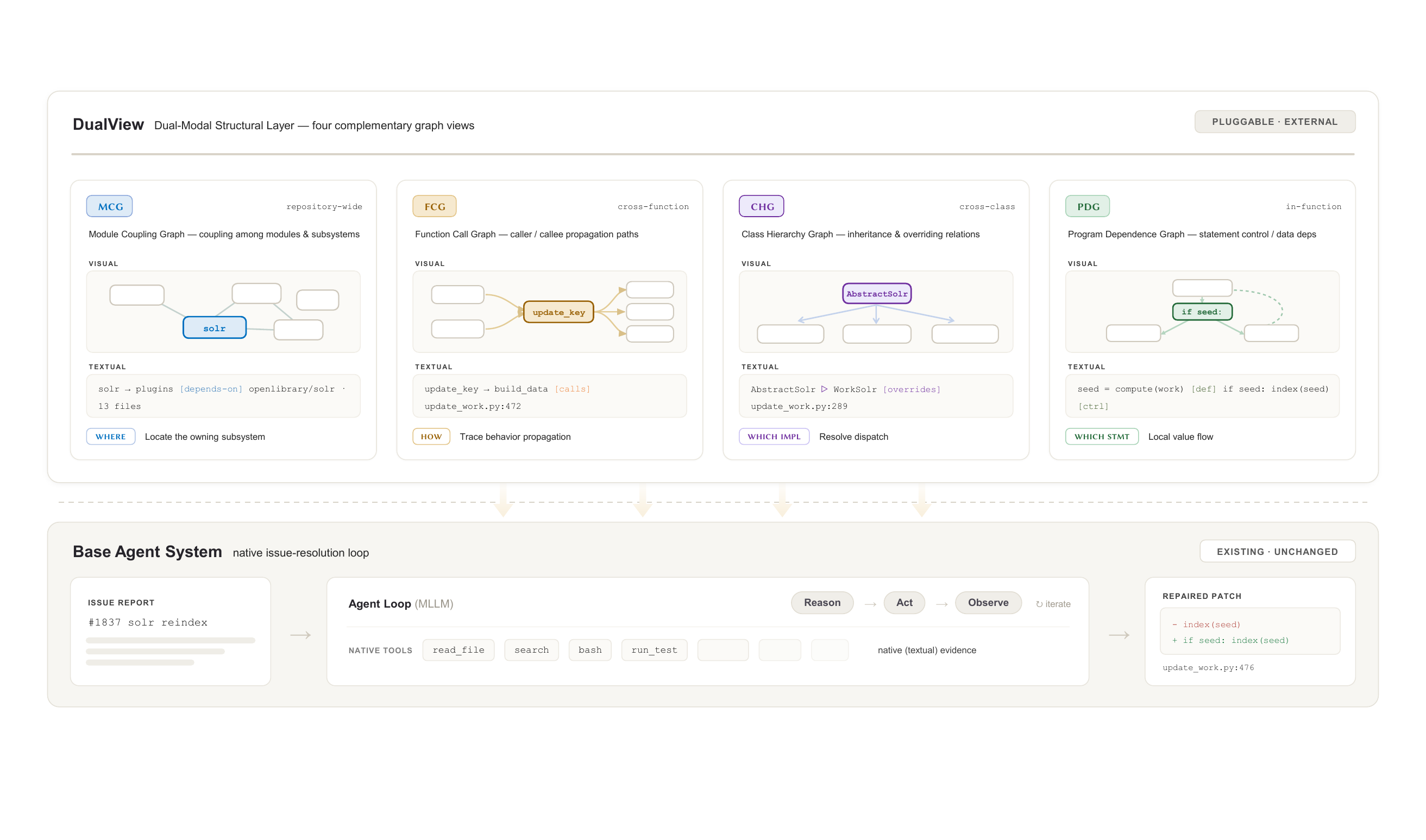}
    \caption{Overview of \TOOL. \chen{color theme is too light, cannot see the text clearly.}}
    \label{fig:overview}
    \vspace{-6pt}
\end{figure*}

% \kai{我建议加一下我们提供该方法的动机，就是因为现有的代码库探索方式与代码库原生的组织方式不匹配（碎片化的文本化观察 vs 结构化的图依赖关系），从而导致了现有的agent系统不准确甚至低效的代码库探索从而限制了现有agent的修复能力。因此我们提出了我们的方法通过利用多个图views来暴露文本化和视觉化的图依赖关系，从而高效帮助agent系统实施高效精确的代码库探索。}
% This section presents \TOOL, a dual-modal structural scaffolding framework for repository-level issue resolution. Figure~\ref{fig:overview} illustrates the overall workflow.
% Given an issue report, the agent iteratively explores the repository, collects evidence, localizes the root cause, and synthesizes a repair. Unlike existing issue-resolution agents that primarily rely on fragmented textual observations, \TOOL exposes repository structure as queryable structural artifacts. Specifically, \TOOL organizes repository relationships into four complementary software-engineering graphs and presents them through both visual and textual modalities. These structural observations provide explicit evidence about repository organization, dependency propagation, inheritance relations, and program dependencies, enabling more effective repository exploration and long-horizon reasoning.

% \kai{我补充一下dualview在应用到现有的agent系统中的pipeline。首先，在agent系统接收到一个issue report后，它会执行分析并通过探索代码库来理解缺陷根因，定位缺陷位置，生成修复补丁等等。在这一过程中，探索代码库需要可以结合DualView提供的4个不同粒度的代码依赖图查询工具来快速重建和呈现代码关系线索，从而实施精确导航等等...你可以根据情况来补充一些具体的pipeline}

\kai{This section presents \TOOL, a dual-modal structural scaffolding framework for repository-level issue resolution. Figure~\ref{fig:overview} provides an overview of the workflow.
% Repository-level issue resolution requires agents to repeatedly explore the repository in order to understand the reported behavior, localize the root cause, collect repair-relevant evidence, and validate candidate patches. 
Repository-level issue resolution requires iterative repository exploration to collect structural evidence and source-level information for bug localization and repair.
% Existing agents primarily perform this exploration through fragmented textual observations, requiring repository structure to be reconstructed incrementally during reasoning.
Existing agents perform this exploration primarily through textual observations, forcing repository structure to be reconstructed incrementally during reasoning.
% \TOOL augments this exploration process by making repository structure directly accessible as explicit structural observations.
\TOOL augments this process by exposing repository structure as explicit dual-modal structural observations.
Specifically, whenever the agent requires structural evidence during repository exploration, it invokes one of the graph-query tools provided by \TOOL. 
% Each query extracts a task-specific structural view from the repository, constructs the corresponding graph, and returns synchronized visual and textual representations derived from the same graph slice. 
Each query extracts a task-specific graph view and returns synchronized visual and textual representations derived from the same graph slice.
The agent combines these dual-modal structural observations with native agent tools to navigate the codebase, inspect implementation details, and progressively localize and repair the bug.
The remainder of this section first introduces the graph-view construction, then presents the dual-modal structural representations, and finally describes how \TOOL is integrated into existing issue-resolution agents.}

\begin{table*}[t]
\centering
\caption{Graph views and parameterized query interfaces exposed to the agent by \TOOL.}
\label{tab:structural_views}
\footnotesize % 稍微缩小一点字号以适应5列
\renewcommand{\arraystretch}{1.2}
\begin{tabular}{p{0.04\linewidth} p{0.16\linewidth} p{0.23\linewidth} p{0.22\linewidth} p{0.23\linewidth}}
\toprule
\textbf{View} & \textbf{Graph Node $v\in V$} & \textbf{Graph Edge $e\in E$} & \textbf{Agent Query Interface $\mathcal{V}(t,s,\theta)$} & \textbf{Structural Evidence} \\
\midrule
\textbf{MCG} & Modules, directories, or package subsystems & Aggregated file-level imports and coupling with direction and weights & $\mathcal{V}(\mcg,\varnothing,\kappa)$: no seed; \newline$\kappa$ caps subsystems & Repository organization and neighboring coupled modules \\
\textbf{FCG} & Functions and methods & Caller--callee invocations & $\mathcal{V}(\fcg,f,\langle h,\delta\rangle)$;\newline\ \
$\delta\!\in\!\{\textit{callers},\textit{callees},\textit{both}\}$ & Behavioral propagation across functions and files \\
\textbf{CHG} & Classes, interfaces, and abstract types & Inheritance, implementation, and overriding & $\mathcal{V}(\chg,C,\langle h,\delta\rangle)$;\newline\ \
$\delta\!\in\!\{\textit{upstream},\textit{downstream},\textit{both}\}$ & Implementation ownership under object-oriented dispatch \\
\textbf{PDG} & Statements, predicates, and variable operations & Data and control flow dependencies & $\mathcal{V}(\pdg,f,\alpha)$;\newline\ \ $\alpha$: optional focus variable / line & Local value flow, branching guards, and state updates \\
\bottomrule
\end{tabular}
\vspace{-6pt}
\end{table*}

\subsection{Multi-Grained Structural Views}

\kai{Repository-level issue resolution progressively narrows the search space from repository-level organization to concrete program statements, requiring structural evidence at different levels of abstraction~\cite{LLM_APR_survey,Agentic_APR_Survey}. To support this process, \TOOL organizes repository structure into four complementary graph views spanning multiple granularities. These views capture four recurring classes of structural relationships encountered during repository exploration: module coupling, function invocation, class inheritance, and statement-level data/control dependence.}
Together, the four views provide a coarse-to-fine exploration hierarchy. Agents first identify relevant subsystems, then analyze interprocedural interactions, object-oriented relationships, and finally fine-grained dependencies around candidate implementations.
Table~\ref{tab:structural_views} summarizes the four structural views provided by \TOOL.

To provide a unified interface across views, \TOOL exposes them through a common view-query abstraction.
Each view is represented as a typed directed graph $G=(V,E)$ and is materialized on demand through a view query:
\begin{equation}
G=\mathcal{V}(t,s,\theta),\quad
t\in\{\mcg,\fcg,\chg,\pdg\}
\label{eq:view_query}
\end{equation}
where $t$ specifies the graph view, $s$ is the seed entity (a function, class, or $\varnothing$ for the repository-level view), and $\theta=\langle h,\delta,\alpha,\kappa\rangle$ controls the query scope, including hop radius, expansion direction, optional intra-view focus, and node budget. The agent issues the query through a graph tool, which constructs $G$ from a precomputed code index and returns the rendered view $R(G)$.

\subsubsection{\textbf{Module Coupling Graph}}
To begin this top-down exploration, the agent must first navigate the overarching structure of the software system. Inspired by the concepts of modular programming~\cite{Module_Organization,Modules,Modules_in_SE}, modern repositories group functionally related code into distinct modules that interact through clear dependencies. Because of this design, the first logical step in narrowing the search space is identifying \emph{which} specific module or subsystem is responsible for the reported behavior. To provide structural evidence for this step, we introduce the \textbf{Module Coupling Graph (\mcg)}~\cite{Module_Coupling}.

\textit{a) Construction.} 
The \mcg is a weighted directed graph
$G_{\mcg}=(V_m,E_c)$. Each node $v\in V_m$ is a \emph{subsystem}, obtained by
partitioning the main source package into its top-level directories (descending
through single-child package nests until a directory exposes at least three
source subdirectories). Every subsystem carries a role tag $\rho(v)$ assigned by deterministic path-pattern rules over the directory leaf name
(e.g.\ \textit{core}, \textit{io\_boundary}, \textit{tests}), a file count, a
natural-language description mined from its README/package docstring, and its
most-imported representative files. A coupling edge $(A,B,w)\in E_c$ is induced
by cross-subsystem imports, where the weight
$w(A,B)=\lvert\{(f_a,f_b): f_a\!\in\!A,\, f_b\!\in\!B,\, f_a\text{ imports }f_b\}\rvert$
counts import references from a file in $A$ to a file in $B$; an edge is retained
only when $w\ge\tau_c$ (e.g., $\tau_c{=}5$) and ranked by weight. This yields a
compact repository-level map in which edge thickness encodes coupling strength.

\textit{b) Query interface.} 
% The agent issues $\mathcal{V}(\mcg,\varnothing,\theta)$ as $\mathcal{V}(\mcg,\varnothing,\kappa)$, 
% with no seed; $\kappa$ caps the number of subsystems. The view is the standard
% entry point on an unfamiliar repository.
The agent issues $\mathcal{V}(\mcg,\varnothing,\kappa)$, where $\kappa$ limits the number of returned subsystems. This view serves as the default entry point for exploring a repository.

\textit{c) Why useful.}
\mcg narrows the search space from the entire repository to a few candidate subsystems while preserving neighboring modules that may also contribute to the issue.
% The \mcg maps out these modules and their coupling relationships. This view gives the agent a clear repository-level orientation, effectively reducing the search space from the entire codebase down to a few candidate subsystems, while still keeping track of neighboring modules that might be involved in the issue.

\subsubsection{\textbf{Function Call Graph}}
% \paragraph{Motivation.} Once a candidate subsystem is identified, exploration
% shifts to concrete entities. Reading a function in isolation does not reveal
% where it is invoked or how behavior propagates across files.
Once the agent identifies a candidate subsystem, the exploration shifts downward to concrete program entities. However, simply finding a relevant function or method is rarely enough; reading its body in isolation does not reveal where it is called from or how it interacts with the rest of the system. To address this, the \textbf{Function Call Graph (FCG)}~\cite{Call_Graph} exposes the caller-callee paths.

\textit{a) Construction.}
The \fcg is a directed graph $G_{\fcg}=(V_{f},E_{\text{call}})$
whose nodes are functions and methods (each with its file and line span) and
whose edges $u\!\to\!v$ denote that $u$ may call $v$, weighted by a static
resolution confidence $c(u,v)\in[0,1]$. Given a seed function $s$, \TOOL
materializes the \emph{ego-graph} by bounded breadth-first expansion to radius
$h$ in direction $\delta\in\{\textit{callers},\textit{callees},\textit{both}\}$;
expansion is direction-stable (callers continue upward, callees downward) to
avoid combinatorial blow-up through hub utilities. Edges below a confidence
floor are discarded, and when $|V_{f}|>\kappa$ the least
important nodes are collapsed into a summary node that reports the count of
hidden callers/callees.

\textit{b) Query interface.} 
The agent issues $\mathcal{V}(\fcg,s,\theta)$ as $\mathcal{V}(\fcg,f,\langle h,\delta\rangle)$, where $f$ is a function (optionally disambiguated by a file hint), $h$ default $2$, and
$\delta$ default \textit{both}.

\textit{c) Why useful.} 
\fcg exposes the caller--callee context around a suspect function, enabling the agent to trace behavior propagation across functions and files before editing.
% \fcg exposes the caller--callee paths around a suspect
% function---its change \emph{blast radius}---letting the agent trace how data and
% behavior cross function and file boundaries before editing.

\subsubsection{\textbf{Class Hierarchy Graph}}
% \paragraph{Motivation.} In object-oriented repositories an issue clue may point
% to an interface, base class, or abstract method, while the faulty logic lives
% in a specific subclass.
% In object-oriented repositories, agents face another layer of uncertainty: abstraction. A clue from an issue report might point to a high-level interface, a base class, or an abstract method, but the actual faulty logic is often hidden in a specific subclass. The \textbf{Class Hierarchy Graph (CHG)}~\cite{Class_Hierarchy} captures this inheritance and overriding structure. 
In object-oriented repositories, issue reports often reference an interface, base class, or abstract method, while the faulty implementation resides in a concrete subclass. To bridge this abstraction gap, we introduce the \textbf{Class Hierarchy Graph (CHG)}~\cite{Class_Hierarchy}.

\textit{a) Construction.} 
\jy{The \chg is a directed graph $G_{\chg}=(V_{t},E_{h})$ where nodes are user-defined types (e.g., \textit{classes}, \textit{interfaces}, \textit{structs}) and edges $child\!\to\!parent$ denote generalized subtype or conformance relations. Because abstraction mechanisms vary across languages, \TOOL constructs $E_h$ on a best-effort static basis. Explicit declarations (e.g., \texttt{extends}/\texttt{implements}) are parsed directly from the AST. Conversely, implicit relationships (e.g., Go interfaces) are inferred by statically matching their declared methods. From a seed type $C$, \TOOL performs bounded BFS to radius $h$ (e.g., $1\!\le\!h\!\le\!5$) in the requested direction, capping wide fan-outs with a summary node to maintain legibility.}
% The \chg is a directed graph $G_{\chg}=(V_{t},E_{h})$ whose
% nodes are types---\textit{class} or \textit{interface}---and whose edges
% $child\!\to\!parent$ carry a relation $r\in\{\textit{extends},\textit{implements}\}$.
% From a seed class $s$, \TOOL performs bounded BFS to radius $h$
% (e.g., $1\!\le\!h\!\le\!5$) in direction $\delta\in\{\textit{upstream}\,(\text{ancestors}),
% \textit{downstream}\,(\text{descendants}),\textit{both}\}$, capping wide fan-outs
% with a summary node so that hot base classes remain legible.

\textit{b) Query interface.} 
The agent issues $\mathcal{V}(\chg,s,\theta)$ as $\mathcal{V}(\chg,C,\langle h,\delta\rangle)$, where $C$ is a class name and \jy{$\delta$ dictates the trace direction (\textit{upstream}, \textit{downstream}, or \textit{both}).} %$\delta$ selecting ancestors, descendants, or both.

\textit{c) Why useful.} 
\chg bridges abstract API clues and their concrete implementations, enabling the agent to identify the actual implementation that should be inspected or edited. Together with \fcg, it complements behavioral relationships with inheritance structure.
% \chg bridges an abstract API clue and the concrete
% implementation to inspect. Together, \fcg and \chg refine candidate clues into
% actionable relationships---execution (invocation) and structure (inheritance).

\subsubsection{\textbf{Program Dependence Graph}}
% \paragraph{Motivation.} 
% Once the agent reaches a candidate implementation, the
% search space narrows to one function body; the remaining question is \emph{which}
% predicate or value flow drives the faulty behavior.
% The \textbf{Program Dependence Graph (PDG)}~\cite{Program_Dependence} supports this final step by exposing statement-level control and data dependencies within a function.
Once the agent reaches a candidate implementation, the remaining task is to identify which control decision or value flow causes the faulty behavior. To support this fine-grained reasoning, we introduce the \textbf{Program Dependence Graph (PDG)}~\cite{Program_Dependence}.

\textit{a) Construction.} 
The \pdg is built for a single function as
$G_{\pdg}=(V_{s},E_{d}\cup E_{ctl})$. Each node $v\in V_{s}$ is a top-level
statement, annotated with the variables it \emph{defines} and \emph{uses} and
its source line. A branch, loop,
or \texttt{try} together with its body forms one node. A data-dependence edge $E_d$ connects a definition to a later use of the same variable, where each use is linked to its nearest preceding definition, and the edge is labeled with the carried variable. A conditional edge $E_{ctl}$ connects a compound predicate (\texttt{if}/\texttt{for}/\texttt{while}) to each later statement that uses a name bound inside its body.
Statements are extracted from the AST by per-language front-ends (Python \texttt{ast}; tree-sitter for Go, JavaScript, and TypeScript) and emitted through a shared assembler, so the dependence analysis is
identical across languages. This analysis is lightweight, focusing on the dominant data and control paths rather than exhaustive intra-procedural coverage. An optional focus $\alpha$ slices the graph to the $h$-hop neighborhood
of a variable or line, and the result is capped at $\kappa$ statements by retaining the most-connected dependency core.

\textit{b) Query interface.} The agent issues $\mathcal{V}(\pdg,s,\theta)$ as $\mathcal{V}(\pdg,f,\alpha)$,
% with
% $f$ a function and, optionally, $\alpha$ to trace one variable's flow.
% The agent issues $\mathcal{V}(\pdg,f,\langle\alpha\rangle)$, 
where $f$ denotes the seed function and $\alpha$ optionally specifies a variable or source line of interest.

\textit{c) Why useful.} 
\pdg exposes statement-level data and control dependencies, enabling the agent to identify the value flow or control decision responsible for a defect before editing.
% \pdg surfaces statement-level control and data
% dependencies, enabling fine-grained reasoning about the exact value flow or
% branch responsible for a defect before the agent commits to an edit.

Together, these four views follow the narrowing logic of repository exploration: MCG guides repository-level orientation, FCG and CHG connect candidate symbols to relevant implementations, and PDG focuses reasoning on statement-level dependencies. Importantly, this top-down progression reflects the inherent structural hierarchy of the software, rather than a mandatory execution pipeline. Depending on the initial issue report and intermediate findings, the agent can flexibly enter the exploration at any level and dynamically switch between views as needed. 

\chen{Why these 4 graphs? Please add supporting evidence.}

\chen{Please give the definition of each graph, and more explanation.}

\subsection{Dual-Modal Structural Observations}

% \kai{Modern MLLMs can jointly reason over visual and textual inputs, yet the two modalities exhibit different representational strengths for structural reasoning~\cite{wang2024picture,Visual_Sketchpad}. Rather than treating one modality as a replacement for the other, \TOOL intentionally separates \emph{structural perception} from \emph{structural grounding} by exposing each queried repository graph through paired visual and textual representations.}
% \jy{Both representations are derived from the same graph slice and therefore preserve identical node and edge structure. They differ in surface detail by design: the visual modality abbreviates node labels to maintain layout legibility and emphasize topology perception, while the textual modality retains full source paths and edge semantics for precise navigation.}

Modern MLLMs can jointly reason over visual and textual inputs, yet the two modalities offer complementary strengths for structural reasoning~\cite{wang2024picture,Visual_Sketchpad}. Rather than relying on a single representation, \TOOL exposes every queried repository graph through synchronized visual and textual observations.
Both observations are derived from the same graph slice and therefore represent the same structural context. The visual modality emphasizes topology perception, whereas the textual modality provides the semantic grounding required for precise repository navigation and repair. The following subsections describe the two representations in detail.

\subsubsection{Visual Structural Representation}

% The visual representation aims to preserve the structural topology of repository relationships rather than maximizing semantic density.
% By rendering graph slices as node-link diagrams, the overall organization, connectivity, and dependency patterns become directly observable, preserving the multi-hop topology that is typically flattened by textual serialization.

% However, embedding complete semantic information (e.g., full file paths, dependency attributes, or implementation details) directly into graph nodes would substantially enlarge the visualization, obscure the underlying topology, and reduce readability. Therefore, \TOOL intentionally keeps each visual node lightweight by displaying only concise identifiers (e.g., symbol name, file name, and line number), while relocating detailed semantic information to the synchronized textual representation. This separation allows the visual modality to emphasize structural perception while preserving a clean and compact layout.

% Although visual and textual representations are generated from the same graph slice, the visual layout exposes several structural cues that are difficult to perceive from sequential text alone:

The visual representation prioritizes preserving repository topology rather than maximizing semantic density. By rendering each graph slice as a node-link diagram, structural organization, connectivity, and dependency patterns become directly observable, preserving relationships that are otherwise flattened by textual serialization.
To maintain a compact and readable layout, \TOOL displays only lightweight node labels (e.g., symbol name, file name, and line number) while relocating complete semantic information (e.g., full paths and dependency attributes) to the synchronized textual representation. This design allows the visual modality to emphasize structural perception without cluttering the graph.
Although both modalities encode the same graph, the visual layout makes several structural cues immediately perceptible:

\begin{itemize}[leftmargin=*]

\item \textbf{Global Organization.}
Spatial grouping reveals subsystem boundaries, clusters, and highly coupled regions, helping agents quickly orient themselves within the repository. This cue is particularly useful for the \mcg.

\item \textbf{Dependency Topology.}
Node-link layouts expose multi-hop propagation paths, fan-in/fan-out patterns, bridge nodes, and densely connected regions without reconstructing connectivity from sequential text. This property is most evident in the \fcg\ and \pdg.

\item \textbf{Structural Alternatives.}
Visual layouts present inheritance hierarchies, alternative callees, and control-flow branches simultaneously, facilitating direct comparison of competing implementations or execution paths. This cue is particularly useful for the \chg\ and \pdg.

\end{itemize}

Overall, the visual modality enables agents to perceive repository topology directly, providing structural guidance that complements conventional text-based repository exploration.

\subsubsection{Textual Structural Representation}

% The textual representation complements the visual modality by preserving semantic information that cannot be compactly encoded in graph layouts. While visual graphs excel at exposing structural topology, they intentionally omit verbose information to maintain readability. \TOOL therefore serializes the same graph slice into a structured textual representation that grounds every visual element in the underlying repository. The textual representation emphasizes four categories of source-grounded information:

The textual representation complements the visual modality by preserving semantic information that cannot be compactly encoded in graph layouts. While visual observations emphasize topology, \TOOL serializes the same graph slice into a structured representation that grounds every visual node and edge in the underlying repository. The textual representation records four categories of source-grounded information:

\begin{itemize}[leftmargin=*]

\item \textbf{Entity Identity.}
Each graph node is mapped to its exact repository entity, allowing symbols with identical display names to be distinguished unambiguously. This information is particularly important for the \fcg\ and \chg, where overloaded methods, inherited implementations, or repeated class names frequently appear across different files.

% \item \textbf{Source Grounding.}
% Every entity is accompanied by its complete repository path and source location, enabling the agent to directly navigate from a structural observation to the corresponding implementation. This grounding mechanism is shared across all four structural views, allowing observations at different granularities to be traced back to concrete source code.

\item \textbf{Source Grounding.}
Each entity is associated with its file path and code location, enabling the agent to directly inspect the corresponding implementation. This grounding is shared across all graph views, allowing structural observations at different granularities to be traced back to source code.

% \item \textbf{Relation Semantics.}
% Graph relationships are serialized into structured dependency statements that explicitly preserve the semantics represented in each structural view, including subsystem coupling (\mcg), function invocations (\fcg), inheritance relationships (\chg), and statement-level data/control dependencies (\pdg).

\item \textbf{Relation Semantics.}
Each graph relation is serialized as a structured dependency statement, preserving the semantics of module coupling (\mcg), function invocation (\fcg), inheritance (\chg), and data/control dependence (\pdg).

\item \textbf{Query Context.}
Additional metadata, including traversal direction, expansion depth, and graph statistics, record how the graph slice was extracted. These contextual cues help the agent interpret the scope of each queried structural view.

\end{itemize}

% Overall, the textual modality complements the visual representation by grounding each structural observation in concrete repository entities. Across the four graph views, it enables the agent to translate high-level structural evidence into executable repository actions, including locating implementations, inspecting dependency relations, and editing the correct source code.

Overall, the textual modality grounds structural observations in concrete repository entities, enabling the agent to seamlessly translate high-level structural evidence into source-level actions such as locating implementations, inspecting dependencies, and editing the correct code.

\begin{figure}[t]
    \centering
    \vspace{2pt}
    \includegraphics[width=\linewidth, trim=19 90 19 90, clip]{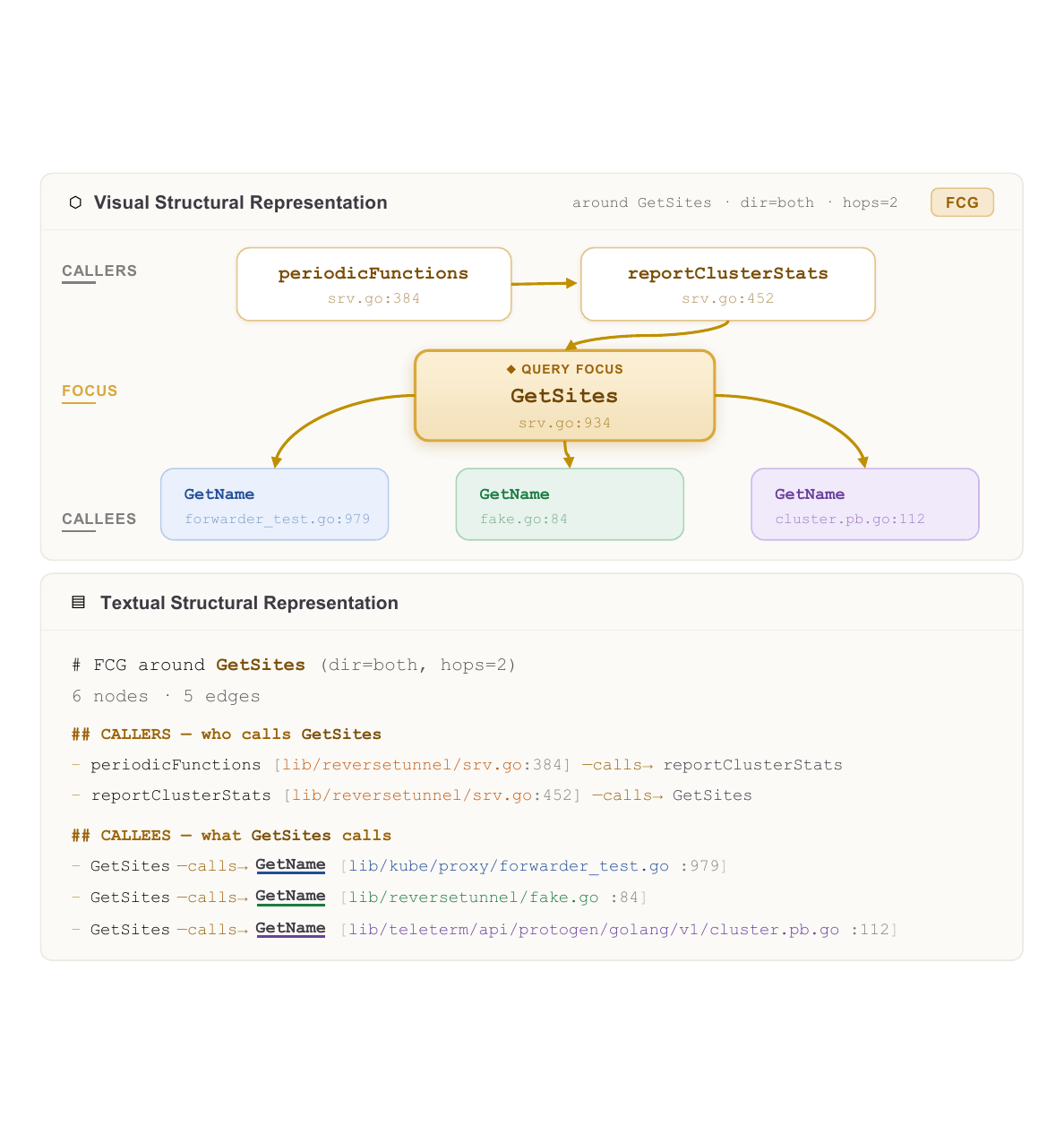}
    \caption{Dual-modal observation of \TOOL.}
    \label{fig:fcg_graph_view_example}
    \vspace{-8pt}
\end{figure}

% \begin{wrapfigure}{r}{0.24\textwidth} 
%     \centering
%     \vspace{-13pt}
%     \includegraphics[width=\linewidth]{Figures/graph_view_example.pdf}
%     % \vspace{-14pt}
%     \caption{Example dual-modal graph observation returned by an \fcg query.}
%     \label{fig:fcg_graph_view_example}
%     % \vspace{-14pt}
% \end{wrapfigure}

% Figure~\ref{fig:fcg_graph_view_example} illustrates an example of how visual and textual representations describe an FCG query around \texttt{GetSites}. 
% \kai{Figure~\ref{fig:fcg_graph_view_example} shows an example dual-modal observation returned from the same FCG query.}
\subsubsection{Dual-Modal Observation in Practice}
Figure~\ref{fig:fcg_graph_view_example} illustrates the dual-modal response 
returned for the same \fcg\ query. 
% from the same FCG query.
Both the visual graph and the textual serialization are generated from an identical graph slice, ensuring that they expose the same structural information while emphasizing complementary aspects of repository reasoning.
In the visual structural representation, the agent can immediately observe a caller chain from \texttt{periodicFunctions} to \texttt{reportClusterStats} and then to \texttt{GetSites}, as well as a downstream fan-out from \texttt{GetSites} to three \texttt{GetName} callees. This structure suggests that the behavior around \texttt{GetSites} may need to be understood both from its upstream use and from its downstream helper calls.
The textual structural representation grounds the same structure in concrete source locations. It separates callers from callees, records each relation as a calls edge, and provides full paths and line numbers.
% for every node. 
Specifically for the three \texttt{GetName} callees, which share the same short name but are distinguished by the file name as the unique identifier. Thus, the agent can use the image to identify the relevant dependency pattern and then use the text to inspect the exact files and lines involved.

\subsection{Adaptive Structural Reasoning}

% Switches from 1) Native / Strctural Evidence 2) If structural evidence --then--> select a approapriate structrual view to collect the relevant evidence

\begin{figure}[t]
    \centering
    \vspace{4pt}
    \includegraphics[width=\linewidth, trim=0 0 0 2, clip]{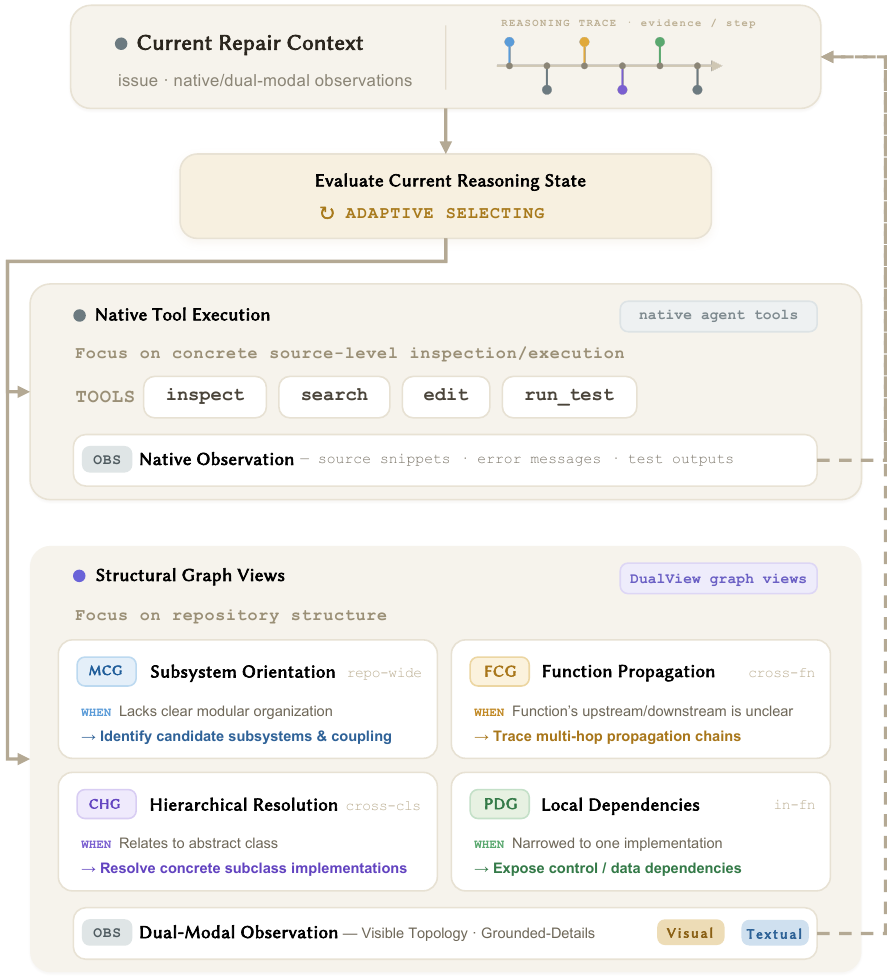}
    \caption{Adaptive reasoning procedure of \TOOL.}
    \label{fig:reasoning}
    \vspace{-8pt}
\end{figure}

To make effective use of the multi-grained structural views and dual-modal structural observation, we propose an adaptive structural reasoning layer inside the agent's original repair loop. As illustrated in Figure~\ref{fig:reasoning}, issue resolution can be considered as a process of gradually turning an issue report into repair actions by collecting a series of evidence within the repair contexts~\cite{SWEBench,zhang2024autocoderover}. 
Existing agents rely exclusively on \textbf{Native Tool Execution} (e.g., source inspection, search) to focus on concrete source-level facts. \TOOL expands the agent's action space by exposing four \textbf{Structural Graph Views} to satisfy the need for repository structure.

% \kai{During issue resolution, the agent incrementally collects evidence, updates its hypotheses, and decides the next action. Existing agents rely exclusively on native repository tools such as search, inspect, and edit. Consequently, whenever structural context is missing, the agent has to reconstruct repository relationships through repeated textual exploration.
% Rather than replacing the original workflow, \TOOL augments this reasoning loop with structural observations.}

% Figure~\ref{fig:reasoning} illustrates how \TOOL integrates structural observations into an existing issue-resolution agent. Rather than replacing the original repair workflow, \TOOL augments the agent's reasoning loop with structural graph queries that are invoked only when repository-level structural evidence is required.

% During issue resolution, the agent incrementally collects evidence, updates its repair hypothesis, and decides the next action. Conventional agents rely entirely on native repository tools (e.g., \texttt{search}, \texttt{inspect}, and \texttt{edit}), forcing them to reconstruct repository structure through repeated textual exploration. \TOOL augments this workflow by introducing structural graph queries that provide explicit repository context whenever structural evidence becomes necessary.

Specifically,
the agent continuously evaluates its current reasoning state by analyzing its gathered context. If its reasoning focuses on concrete source-level actions, the agent relies on \textit{Native Tool Execution} (e.g., \texttt{inspect}, \texttt{search}, \texttt{edit}). Conversely, when the agent requires repository structural context to proceed, it transitions to Structural Graph Views. Within this structural track, the following state-driven conditions further guide the agent to invoke the most appropriate graph view:

% A shorter version
\begin{itemize}[leftmargin=*]
\item \textbf{Subsystem Orientation.} At the initial stage of repository exploration, the agent often lacks clear modular organization. To map the issue description to high-level components, the protocol recommends querying the \textbf{\mcg}. This repo-wide view identifies candidate subsystems and coupling relations before fine-grained inspection begins.
\item \textbf{Function Propagation.} If the agent isolates a suspicious function but its upstream or downstream role is unclear, the protocol triggers the \textbf{\fcg}. This cross-function view traces multi-hop propagation chains, exposing caller-callee paths and impact boundaries.
\item \textbf{Hierarchical Resolution.} Clues pointing to an abstract class or interface often indicate that the actual faulty behavior resides elsewhere. To resolve this abstraction, the protocol directs the agent to query the \textbf{\chg}, which explicitly identifies concrete subclass implementations.
\item \textbf{Local Dependencies.} Having narrowed the focus to one specific implementation, the agent must identify the exact faulty state update or predicate. The protocol then suggests the \textbf{\pdg} to expose statement-level control and data dependencies, allowing the agent to reason about variable flows before editing code contents.
\end{itemize}

The reasoning process is therefore adaptive rather than fixed. An agent may begin with the \mcg\ to identify a relevant subsystem, switch to native tools to inspect candidate code, invoke the \fcg\ or \chg\ to refine structural hypotheses, and finally use the \pdg\ for statement-level analysis. Each graph query returns synchronized visual and textual observations, enabling the agent to alternate naturally between structural reasoning and source-level inspection throughout issue resolution.

% This process is adaptive because the agent may switch views as new clues emerge. For example, an \mcg\ observation may direct the agent to a subsystem, source inspection may reveal a candidate function, and an \fcg\ query may then expose a downstream helper that should be inspected. A later \pdg\ query may show that the faulty value depends on a specific branch or helper call, causing the agent to revisit \fcg\ for interprocedural impact analysis. Similarly, an \fcg\ path may reach an abstract method, which then motivates a \chg\ query to identify the concrete implementation. Thus, \TOOL does not enforce a simple coarse-to-fine pipeline. It allows the agent to move between structural views as its repair focus changes.

\subsection{Integration with Existing Agents}
\chen{suggest to move the implementation subsection here.}
To make the adaptive reasoning mechanism applicable to existing repair systems, \TOOL is implemented as an agent-agnostic structural reasoning layer.
Existing agents invoke graph queries through their native tool interfaces. For mini-SWE-agent~\cite{mini_SWE_agent}, graph queries are exposed as shell commands that naturally fit its bash-centric interaction loop. For OpenCode~\cite{OpenCode}, the same functionality is provided through an MCP server. Since both interfaces preserve the original interaction workflow, integrating \TOOL requires only minimal modifications to existing agents.

% This design turns graph views from static context augmentation into on-demand structural evidence. Native textual tools remain appropriate when the missing clue is a local implementation detail or runtime fact. Structural views are invoked when the next repair decision depends on repository relationships. By retrieving only the structural evidence that matches the current exploration objective, \TOOL helps the agent maintain a coherent view of repository structure without injecting unnecessary graph context into every step.

% Tools -> 封装成MCP来支持更多agent
% \subsection{Agent Integration}

% % We implement \TOOL as an external structural-analysis service that can be integrated into existing issue-resolution agents without modifying their underlying reasoning policies.
% We implement DualView as an external structural reasoning layer that augments existing issue-resolution agents without modifying their reasoning policies.

% The framework constructs graph views from repository source code using static analysis and exposes them through reusable tools and MCP services. Since \TOOL operates as a modular structural layer, it can be integrated into different agent frameworks and multimodal language models. In our experiments, we integrate \TOOL into OpenCode and mini-SWE-agent without changing their original repair workflows.
\section{Experiment Setup}
\label{sec:setup}

\subsection{Research Questions}
\label{sec:setup:rq}

% We answer three research questions, each pre-empting a distinct reviewer concern.

\noindent
\textbf{RQ1}: 
\kai{Can \TOOL consistently enhance the repair capability of existing agent systems? (Overall Effectiveness)}
% How does \TOOL compare with state-of-the-art baselines on long-horizon task instances?
\chen{As far as I understand, what you develop is just a MCP tool/plugin, and how do you compare with SOTA agent? Maybe "Is \TOOL effective in bug repair?"}
% This question evaluates whether visual scaffolding produces a measurable end-to-end gain over the text-only agent systems on the long-horizon task instances under a fixed base model.

% \noindent
% \textbf{RQ2 (Ablation Study): How do different graph views and the visual modality contribute to \TOOL's effectiveness?}
% This question slices the gain along two orthogonal axes. First, it removes one graph view at a time to expose each view's marginal contribution; second, it uses equivalent text-only graph descriptions to isolate the contribution of the visual modality itself.

\noindent
\textbf{RQ2}: 
\kai{How do multi-grained views and dual-modal observations contribute to \TOOL? (Ablation Study)}
% How do repository-structure modeling and visual reasoning individually contribute to \TOOL?
% This question disentangles two key factors underlying \TOOL. First, we evaluate the contribution of each graph view by removing it from the framework and measuring the resulting performance degradation. Second, we replace visual graph renderings with equivalent textual descriptions to determine whether the gains stem solely from exposing structural information or from presenting that information in a visual form that better supports long-horizon reasoning.

\noindent
\textbf{RQ3}: How well does \TOOL generalize to additional issue resolution benchmarks?  (Generalizability Study)
% This question evaluates whether the gain is bound to a single benchmark by re-running on \swebv.
% This question investigates whether the effectiveness of \TOOL arises from benchmark-specific characteristics or from a more general capability to support visual structural reasoning. To evaluate its robustness, we replicate experiments on \swebv .

% \subsection{Benchmarks}
% \label{sec:setup:benchmarks}

% \noindent
% \textbf{\swebp} is the primary benchmark for RQ1 and RQ2. It's public set contains 731 long-horizon task instances drawn from real GitHub repositories, hand-curated for solvability and instrumented with hidden test suites. We use the full Public set for RQ1 and a 50-instance subset for RQ2. Note that the same subset Live-SWE-agent~\cite{Live_Swe_Agent} reports on, for the seven-variant ablation in RQ2.
% \kai{注意我们选择在RQ2消融实验仅仅选择50个instances的子集上进行是考虑到消融实验巨大的实验成本，我们设计了7个变体来澄清不同组件设计的贡献。这样的做法是合理的，因为我们follow的Live-SWE-agent~\cite{Live_Swe_Agent}的做法，它的消融实验也是选择了50-instance来实施的。}
% \jy{我把这里改了，50个instance的子集的消融效果不明显，提高测试数量后变体之间的差异才得已体现}

% \textbf{\swebv} is the secondary benchmark for RQ3~\cite{jimenez2023swe,openai2024swev}. It contains 500 instances drawn from earlier GitHub history. We use it to verify the generalizability of the scaffolding across benchmarks, because Pro and Verified differ in repository selection with various program languages, issue type distribution, and patch difficulty.

\subsection{Benchmarks}
\label{sec:setup:benchmarks}

\noindent
\textbf{\swebp}~\cite{SWEBench_Pro} is used for RQ1 and RQ2. Its Public split contains 731 long-horizon issue-resolution instances. 
% We use the full Public set for RQ1 to assess the overall effectiveness.
Note that we conduct ablation studies on a 150-instance randomly selected subset of \swebp. Evaluating multiple variants on the full benchmark would incur substantially higher costs, while the subset provides a widely adopted setting for analyzing the contribution of individual components.

\noindent
\textbf{\swebv}~\cite{SWEBenchVerfied} is used for RQ3 to evaluate generalizability. It contains 500 task instances collected from real GitHub repositories and has been widely adopted to assess the repair capabilities of repair agents or base models. We use it to examine whether the benefits of visual structural reasoning transfer across benchmarks with different repository characteristics, issue distributions, and task compositions.

\subsection{Baselines}
\label{sec:setup:baselines}
We evaluated \TOOL against two categories of baselines to systematically isolate our identified limitations. 
First, we select four representative text-centric agent systems: SWE-agent~\cite{SWE_Agent}, mini-SWE-agent~\cite{mini_SWE_agent}, Live-SWE-agent~\cite{Live_Swe_Agent}, and OpenCode~\cite{OpenCode}. 
All of them rely primarily on text-based repository exploration.
Second, to compare 
% the usefulness of textual graph-structured repository representation, 
the closest textual graph-structured repository representation approaches,
we select RepoGraph~\cite{RepoGraph} and CodeGraph~\cite{CodeGraph}, both of which can be plugged into existing agent systems, making them the most direct baselines for evaluating the effectiveness of our proposed dual-modal structural observations.

% We evaluated \TOOL againsit two categories of baselines to systematically isolate our identified limitations.
% First, 
% % We evaluate four representative issue-resolution agent systems: 
% we select four text-centric agent systems:
% SWE-agent~\cite{SWE_Agent}, mini-SWE-agent~\cite{mini_SWE_agent}, Live-SWE-agent~\cite{Live_Swe_Agent}, and OpenCode~\cite{OpenCode}. All of them rely primarily on text-based repository exploration.
% Second,
% we compare against RepoGraph~\cite{RepoGraph} and CodeGraph~\cite{CodeGraph}, the closest related structural repository representation approaches, which are external plugins that augment repository exploration through graph representations, making them the most direct baselines for evaluating the effectiveness of our dual-modal structural observations.

\subsection{Experimental Configurations}
\label{sec:setup:config}

\chen{Due to the non-determinism, have you run multiple times to get the results in average? If not, may add it to threat to validity.}
Since \TOOL is designed as a repository exploration plugin rather than a standalone issue-resolution agent, our evaluation focuses on measuring the improvement it brings to existing agent systems.
% We therefore integrate \TOOL into agent scaffolds and compare each enhanced version against its corresponding original implementation under identical experimental settings.
We consider two representative open-source systems:
\textbf{mini-SWE-agent}~\cite{mini_SWE_agent}
represents lightweight research-oriented agent systems, and has become one of the most widely adopted foundations for issue-resolution research.
\textbf{OpenCode}~\cite{OpenCode}
represents modern open-source coding agents with a different interaction paradigm and repository exploration workflow.
To evaluate the generality of \TOOL across base models, we instantiate each scaffold using Claude 4.5 Sonnet~\cite{claude45sonnet} and Kimi K2.5~\cite{kimi25} on SWE-bench Pro. For the cross-benchmark study on SWE-bench Verified, we additionally use Gemini 3 Flash~\cite{Gemini3Flash} due to its low-cost multimodal support.

\section{Evaluation}
\label{sec:eval}

% \kai: I provide a new version for each RQ

% \input{5_1_RQ1_old}
\subsection{RQ1: Overall Effectiveness}
\label{sec:eval:rq1}

% \usepackage{multirow}
\begin{table}[]
\vspace{6pt}
\caption{
Improvements of \TOOL with different agent scaffolds and base models on SWE-bench Pro public.
\chen{highlight the number in the last two rows for our approach.}
}
\label{tab:result_RQ1_base_full_improve}
\resizebox{1.0\columnwidth}{!}{
\begin{tabular}{lc|cc|cc|cc}
\toprule
\rowcolor[HTML]{FFFFFF}
\multicolumn{1}{c}{} &
  \multicolumn{1}{c|}{} &
  \multicolumn{2}{c|}{mini-SWE-agent} &
  \multicolumn{2}{c|}{OpenCode} &
  \multicolumn{2}{c}{OpenCode} \\
\rowcolor[HTML]{FFFFFF}
\multicolumn{1}{c}{\textbf{Repo}} &
  \multicolumn{1}{c|}{\textbf{Num}} &
  \multicolumn{2}{c|}{Claude 4.5 Sonnet} &
  \multicolumn{2}{c|}{Claude 4.5 Sonnet} &
  \multicolumn{2}{c}{Kimi K2.5} \\
\rowcolor[HTML]{FFFFFF}
\multicolumn{1}{c}{}               &                & 
\raisebox{-0.20em}{\includegraphics[height=1em]{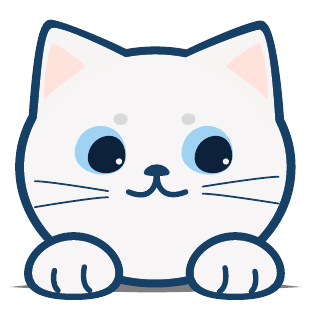}}$_{base}$ & 
\raisebox{-0.20em}{\includegraphics[height=1em]{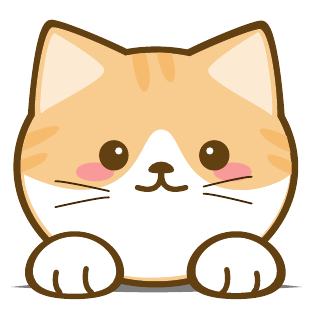}}$_{full}$ & 
\raisebox{-0.20em}{\includegraphics[height=1em]{Icons/base.pdf}}$_{base}$ & 
\raisebox{-0.20em}{\includegraphics[height=1em]{Icons/full.pdf}}$_{full}$ & 
\raisebox{-0.20em}{\includegraphics[height=1em]{Icons/base.pdf}}$_{base}$ & 
\raisebox{-0.20em}{\includegraphics[height=1em]{Icons/full.pdf}}$_{full}$ \\ \midrule
\rowcolor[HTML]{FFFFFF}
ansible (Py)                    & 96              & 51    & 55    & 47    & 55    & 45    & 47    \\
\rowcolor[HTML]{EFEFEF}
openlibrary (Py)                & 91              & 47    & 54    & 41    & 40    & 46    & 52    \\
\rowcolor[HTML]{FFFFFF}
qutebrowser (Py)               & 79              & 49    & 51    & 45    & 47    & 50    & 53    \\
\rowcolor[HTML]{EFEFEF}
teleport (Go)                      & 76              & 29    & 35    & 25    & 28    & 30    & 32    \\
\rowcolor[HTML]{FFFFFF}
flipt (Go)                         & 85              & 30    & 29    & 29    & 31    & 28    & 33    \\
\rowcolor[HTML]{EFEFEF}
vuls (Go)                          & 62              & 33    & 27    & 27    & 32    & 26    & 39    \\
\rowcolor[HTML]{FFFFFF}
navidrome (Go)                     & 57              & 23    & 25    & 22    & 22    & 23    & 25    \\
\rowcolor[HTML]{EFEFEF}
webclients (JS)                    & 65              & 35    & 36    & 28    & 33    & 33    & 36    \\
\rowcolor[HTML]{FFFFFF}
element-web (JS)                   & 56              & 30    & 32    & 18    & 29    & 22    & 30    \\
\rowcolor[HTML]{EFEFEF}
NodeBB (JS)                        & 44              & 17    & 19    & 24    & 24    & 27    & 30    \\
\rowcolor[HTML]{FFFFFF}
tutanota (TS)                      & 20              & 11    & 12    & 10    & 10    & 12    & 11    \\  \midrule
\rowcolor[HTML]{FFFFFF}
\multicolumn{2}{c|}{\textbf{\#Total Resolved (731)}} & 355    & \textbf{375\gain{20}}   & 316    & \textbf{351\gain{35}}    & 342    & \textbf{388\gain{46}}   \\
\rowcolor[HTML]{FFFFFF}
\multicolumn{2}{c|}{\textbf{\$Avg. Cost}}            & \$0.94   & \textbf{\$0.89\save{0.05}}   & \$1.40   & \textbf{\$1.36\save{0.04}}  & \$0.55   & \textbf{\$0.50\save{0.05}}  \\  \bottomrule
\end{tabular}
}
\vspace{0pt}
\end{table}

\subsubsection{Controlled Study}
We first evaluate \TOOL through controlled comparisons in which the agent system and base model are fixed, and the only difference is whether dual-modal scaffolding is enabled. 
As shown in Table~\ref{tab:result_RQ1_base_full_improve}, \raisebox{-0.20em}{\includegraphics[height=1em]{Icons/base.pdf}}$_{base}$ and \raisebox{-0.20em}{\includegraphics[height=1em]{Icons/full.pdf}}$_{full}$ denote the original and \TOOL-enabled agents respectively.
% Here, \raisebox{-0.20em}{\includesvg[height=1em]{Icons/base.svg}}$_{base}$ refers to the performance of the original agent system, while \raisebox{-0.20em}{\includesvg[height=1em]{Icons/full.svg}}$_{full}$ refers to the results after enabling DualView.
% Table~\ref{tab:result_RQ1_base_full_improve} reports the results. 
Across all three settings, integrating \TOOL consistently improves issue-resolution performance. When built on top of \miniswea with Claude 4.5 Sonnet, \TOOL increases the number of resolved instances from 355 to 375 (+20, 5.6\%) while slightly reducing the average cost from \$0.94 to \$0.89. 
% The gains are distributed across most repositories, with improvements observed on 9 of the 11 projects.
The improvements become more pronounced on OpenCode. Under Claude 4.5 Sonnet, \TOOL raises the number of resolved instances from 316 to 351 (+35, 11.1\%), while reducing the average cost from \$1.40 to \$1.36. When replacing the base model with Kimi K2.5, the number of resolved instances further increases from 342 to 388 (+46, 13.5\%), again with a lower average cost (\$0.55 to \$0.50).
These isolated configurations demonstrate that the benefits of \TOOL are consistent across both agent scaffolds and base models.

\subsubsection{Baseline Comparison}

\begin{table}[t]
\caption{
% Resolved instances on \swebp public. 
Comparison with representative agent baselines.
}
\label{tab:rq1}
\centering
\small
\setlength{\tabcolsep}{6pt}
\renewcommand{\arraystretch}{1.15}
\resizebox{1.0\columnwidth}{!}{
\begin{tabular}{llccc}
\toprule
\textbf{Agent System} & \textbf{Base Model} & \textbf{Resolved} & \textbf{\%Resolved} & \textbf{\$Avg.\ Cost} \\
\midrule
% Kimi internal agent~\cite{Kimi_Visual_Agentic_Intelligence}       & \kimi\kimiktwofive                & 371 & 50.7\% & N/A \\
OpenCode       & \kimi\kimiktwofive                & 342 & 46.8\% & \$0.55 \\
OpenCode       & \anthropic\claudesonnetfourfive                & 316 & 43.2\% & \$1.40 \\
% CodeGraph\textsubscript{@OpenCode}       & \anthropic\claudesonnetfourfive                & xx & xx.x\% & \$x.xx \\ 
\midrule
% \miniswea~\cite{SWEBenchPro_leaderboard}       & \openai~GPT-5.4 (xHigh)                & 432 & 59.1\% & N/A \\
% \miniswea~\cite{SWEBenchPro_leaderboard}   & 
% \meta Muse Spark  & 402 & 55.0\% & N/A     \\
% \miniswea~\cite{SWEBenchPro_leaderboard}   & \gemini\geminithreeonepro        & 396 & 54.2\% & N/A \\
% \miniswea~\cite{SWEBenchPro_leaderboard}   & \anthropic\claudeopusfoursix  & 379 & 51.9\% & N/A     \\
\miniswea   & \anthropic\claudesonnetfourfive  & 355 & 48.6\% & \$0.94     \\ 
\liveswea~\cite{Live_Swe_Agent}   & \anthropic\claudesonnetfourfive  & 335 & 45.8\% & \$0.73     \\ 
\swea~\cite{SWEBenchPro_leaderboard}       & \anthropic\claudesonnetfourfive  & 319 & 43.6\% &  N/A \\
% SWE-Adept~\cite{SWE_Adept}       & \anthropic\claudesonnetfourfive  & xxx & 45.0\% &  N/A \\
\midrule
\addlinespace[1pt]
\rowcolor{myLightBlue}
\TOOLat{OpenCode}      & \kimi~Kimi K2.5
    & \textbf{388} & \textbf{53.1\%} & \$0.50 \\
\rowcolor{myLightBlue}
\TOOLat{OpenCode}      & \anthropic\claudesonnetfourfive
    & \textbf{351} & \textbf{48.0\%} & \$1.36 \\
\rowcolor{myLightBlue}
\TOOLat{\miniswea} & \anthropic\claudesonnetfourfive
    & \textbf{375} & \textbf{51.3\%} & \$0.89 \\
\bottomrule
\end{tabular}
}
\end{table}

Table~\ref{tab:rq1} further compares \TOOL with representative issue-resolution agents.
Across both OpenCode and mini-SWE-agent configurations, the \TOOL-enabled agents consistently achieve the best repair performance under identical foundation models, outperforming existing text-centric agent systems.

\begin{figure}[t]
% \vspace{-6pt}
  \centering
  \begin{subfigure}[b]{0.49\linewidth}
    \centering
    \includegraphics[width=\linewidth, trim=0 5 0 7, clip]{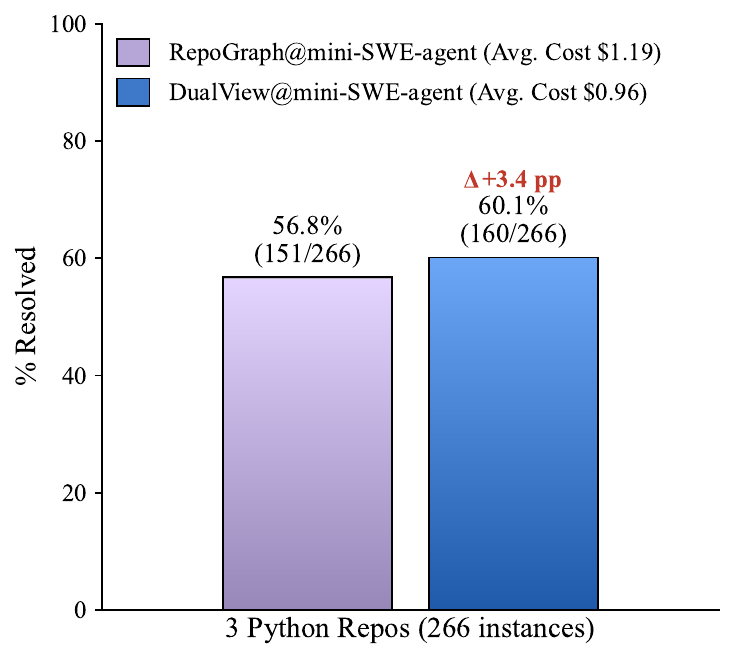}
    \caption{RepoGraph vs. DualView.}
    \label{fig:repograph}
  \end{subfigure}
  \hfill
  \begin{subfigure}[b]{0.49\linewidth}
    \centering
    \includegraphics[width=\linewidth, trim=0 5 0 7, clip]{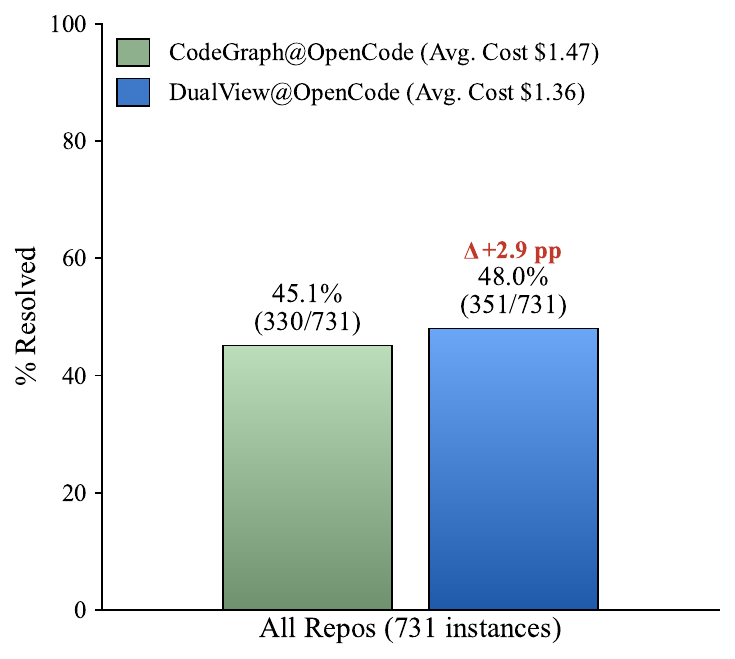}
    \caption{CodeGraph vs. DualView.}
    \label{fig:codegraph}
  \end{subfigure}
  \caption{Comparison with graph repository exploration methods.}
  \label{fig:compare_graph_baseline}
% \vspace{-2pt}
\end{figure}

Beyond comparing with complete agent systems, we compare \TOOL with graph-based repository exploration methods, which are the closest alternatives to our structural scaffolding framework.
RepoGraph~\cite{RepoGraph} represents repository structure as textual graph retrieval and can be integrated into existing agents.
Since it currently supports only Python repositories and is not released as an MCP service, we integrate it into mini-SWE-agent and evaluate both methods on the 266 Python instances in SWE-bench Pro.
We additionally compare with CodeGraph~\cite{CodeGraph}, another graph-based repository exploration system released as an MCP service.
Unlike RepoGraph, CodeGraph natively supports multiple languages via MCP.
% , so we evaluate it with OpenCode on the full benchmark. 
We therefore integrate it into OpenCode and evaluate it on the full SWE-bench Pro benchmark.
Figure~\ref{fig:compare_graph_baseline} summarizes the comparison between \TOOL and 
% \jy{these two} 
two representative graph-based repository exploration
methods with Claude 4.5 Sonnet. \TOOL again achieves superior repair performance and lower cost, demonstrating that the proposed dual-modal multi-view structural observations consistently outperform existing graph-based repository representations.

% \subsubsection{SOTA Comparison}
% To position \TOOL relative to existing issue-resolution systems, Table~\ref{tab:rq1} compares our configurations with recent open-source and commercial agents reported on \swebp.
% Among configurations using Claude 4.5 Sonnet, \TOOLat{\miniswea} resolves 375 instances, outperforming \miniswea (355), Live-SWE-agent (335), and SWE-agent (319) under the same base model. 
% % Notably, it achieves performance comparable to mini-SWE-agent equipped with the stronger Claude 4.6 Opus model (379).
% When integrated with OpenCode, \TOOL improves the resolve rate from 43.2\% to 48.0\% under Claude 4.5 Sonnet, and reaches 53.1\% under Kimi K2.5. The latter configuration surpasses both the original OpenCode implementation and the Kimi internal agent reported by Moonshot AI.
% Overall, these results indicate that visual structural scaffolding can substantially strengthen existing issue-resolution agents and remain competitive with state-of-the-art systems built upon more powerful foundation models.

% \usepackage{multirow}
\begin{table}[t]
\caption{Cost efficiency comparison by \textit{base} and \textit{full}.}
\label{tab:result_RQ1_cost_efficiency}
\resizebox{1.0\columnwidth}{!}{
\begin{tabular}{l|ccc|ccc|ccc}
\toprule
\multicolumn{1}{c}{\multirow{3}{*}{Metrics}} & \multicolumn{3}{|c}{mini-SWE-agent}    & \multicolumn{3}{|c}{OpenCode}          & \multicolumn{3}{|c}{OpenCode}  \\
\multicolumn{1}{c}{}                         & \multicolumn{3}{|c}{Claude 4.5 Sonnet} & \multicolumn{3}{|c}{Claude 4.5 Sonnet} & \multicolumn{3}{|c}{Kimi K2.5} \\
\multicolumn{1}{c}{} &
  \multicolumn{1}{|c}{\raisebox{-0.20em}{\includegraphics[height=1em]{Icons/base.pdf}}$_{base}$} &
  \multicolumn{1}{c}{\raisebox{-0.20em}{\includegraphics[height=1em]{Icons/full.pdf}}$_{full}$} &
  \multicolumn{1}{c}{$\blacktriangle$} &
  \multicolumn{1}{|c}{\raisebox{-0.20em}{\includegraphics[height=1em]{Icons/base.pdf}}$_{base}$} &
  \multicolumn{1}{c}{\raisebox{-0.20em}{\includegraphics[height=1em]{Icons/full.pdf}}$_{full}$} &
  \multicolumn{1}{c}{$\blacktriangle$} &
  \multicolumn{1}{|c}{\raisebox{-0.20em}{\includegraphics[height=1em]{Icons/base.pdf}}$_{base}$} &
  \multicolumn{1}{c}{\raisebox{-0.20em}{\includegraphics[height=1em]{Icons/full.pdf}}$_{full}$} &
  \multicolumn{1}{c}{$\blacktriangle$} \\ 
  \midrule
\rowcolor[HTML]{EFEFEF}
Total resolved                         & 355           & 375          & +20          & 316        & 351        & +35         & 342         & 388       & +46       \\ 
\rowcolor[HTML]{EFEFEF}
Avg. cost / instance                         & \$0.94           & \$0.89          & -\$0.05          & \$1.40     & \$1.36     & -\$0.04     & \$0.55         & \$0.50       & -\$0.05       \\ 
\rowcolor[HTML]{EFEFEF}
Avg. LLM steps / instance                    & 71.2           & 62.7          & -8.5          & 66.0       & 60.0       & -6.0        & 50.1         & 45.5       & -4.6       \\ 
\rowcolor[HTML]{E1E8F2}   % 浅蓝灰
\hspace{0.8em}\textit{— Resolved instances}                & 65.8           & 58.7          & -7.1          & 62.8       & 56.9       & -5.9        & 46.4         & 42.3       & -4.1       \\
\rowcolor[HTML]{E1E8F2}   % 浅蓝灰
\hspace{0.8em}\textit{— Unresolved instances}              & 76.3           & 66.8          & -9.5          & 68.5       & 62.3       & -6.2        & 53.4         & 49.3       & -4.1       \\
\rowcolor[HTML]{EFEFEF}
Avg. graph calls                             & 0.0           & 1.8          & +1.8          & 0.0          & 3.7        & +3.7           & 0.0         & 2.3       & +2.3       \\
\bottomrule
\end{tabular}
}
\vspace{-6pt}
\end{table}

\subsubsection{Cost Efficiency}
\jy{A natural concern is that visual structural scaffolding introduces graph-processing overhead by injecting visual and textual representations. However, Table~\ref{tab:result_RQ1_cost_efficiency} shows that this overhead does not increase overall inference cost. Although \TOOL introduces an average of 1.8 to 3.7 graph calls per instance, the overall number of LLM steps actually decreases, as the structural views prevent invalid exploration and unnecessary context accumulation. For example, OpenCode with Claude 4.5 Sonnet requires fewer LLM steps (dropped from 66.0 to 60.0), and similar reductions are observed for mini-SWE-agent and OpenCode with Kimi K2.5. Consequently, \TOOL successfully offsets this initial overhead, yielding a consistent cost reduction of \$0.04--\$0.05 per instance while simultaneously improving repair performance. This confirms that efficient structural navigation is ultimately more cost-effective than aimless textual search.}
% A natural concern is that visual structural scaffolding introduces graph-processing overhead. However, Table~\ref{tab:result_RQ1_cost_efficiency} shows that this overhead does not increase overall inference cost.
% Although \TOOL introduces an average of 1.8 to 3.7 graph calls per instance, all three configurations require fewer LLM interaction steps than their corresponding baselines. 
% For example, OpenCode with Claude 4.5 Sonnet requires fewer LLM steps (66.0 vs. 60.0), despite an average of 3.7 graph invocations per instance. Similar reductions are observed for mini-SWE-agent and OpenCode with Kimi K2.5.
% As a result, the average cost per instance decreases consistently across all settings, yielding savings of \$0.04--\$0.05 while simultaneously improving repair performance. This suggests that 
% % the structural views provided by 
% \TOOL helps agents search for relevant code regions more efficiently, reducing exploratory interactions and offsetting the additional graph-processing overhead.

% \subsubsection{Summary}
% Across different agent scaffolds and base models,
% \TOOL consistently improves issue-resolution performance
% % (+20 to +46 additional resolved instances on SWE-bench Pro)
% while slightly reducing cost, demonstrating that
% dual-modal structural scaffolding provides a general and cost-effective
% enhancement for repository exploration and repair.

% \input{5_2_RQ2_old}

\subsection{RQ2: Ablation Study}
\label{sec:eval:rq2}

% In this section, we conduct the ablation study to analyze how graph views and visual modality contribute to \TOOL's effectiveness. Based on our analysis in Section~\ref{sec:eval:rq1}, \TOOLat{\opencode} with \kimiktwofive achieves the best overall performance among our evaluated settings while retaining a reasonable overhead, making it a practical choice for evaluating multiple variants. Hence, we design different variants on top of \TOOLat{OpenCode} with \kimiktwofive that progressively isolate the role of each individual graph view and of the visual modality. 

% To understand which components of \TOOL contribute to its effectiveness, we conduct an ablation study on top of \TOOLat{OpenCode} with \kimiktwofive, which achieved the good effectiveness-cost tradeoff in RQ1.
% To control experimental cost, all variants are evaluated on a subset of \swebp. \jy{Notably, the resolution rate on the subset closely matches its rate on the full benchmark, suggesting that the subset is representative.} 
% As shown in Table~\ref{tab:rq2}, these variants allow us to analyze the contribution of each graph view, the superiority of using visual modality, and the complementarity of multiple graph views.
\kai{To understand the contribution of \TOOL's two key designs, multi-grained structural views and dual-modal structural observations, we conduct an ablation study on \TOOLat{OpenCode} with \kimiktwofive, which achieved the best effectiveness-cost tradeoff in RQ1. To control experimental cost, all variants are evaluated on a subset of \swebp. 
As shown in Table~\ref{tab:rq2}, the ablation variants are organized into two groups. The first evaluates the individual contribution of each structural graph view, 
% (\mcg, \fcg, \chg, and \pdg), 
while the second examines the effectiveness of \textit{textual} and \textit{visual} structural observations.}

% To understand the contribution of \TOOL's two key designs, multi-grained structural views and dual-modal structural observations, we conduct an ablation study on \TOOLat{OpenCode} with \kimiktwofive. To reduce computational cost, all variants are evaluated on a representative subset of \swebp. As shown in Table~\ref{tab:rq2}, the ablations evaluate the individual graph views (\mcg, \fcg, \chg, and \pdg) and the two structural observation modalities (textual vs.\ visual).
% \kai{需要解释一下消融实验中为什么用OpenCode而不是mini-swe-agent的原因，以及为什么用Kimi而不是Claude的原因。可以说在RQ1中我们的初步实验发现OpenCode的Kimi实现不仅效果好而且成本低，因此考虑到消融实验需要大规模运行多个变体的成本考量，因此选择了OpenCode的Kimi实现上来验证我们方法设计的贡献}

\begin{itemize}[leftmargin=0.33cm]
\item \raisebox{-0.20em}{\includegraphics[height=1em]{Icons/base.pdf}}~\textbf{\TOOLbase}: We remove all graph views access from \TOOLat{OpenCode} to establish a base version that reflects the fundamental issue-resolution capability of the underlying \llm. This variant is equivalent to original OpenCode.
% against which the contribution of every view is measured.
\item \raisebox{-0.20em}{\includegraphics[height=1em]{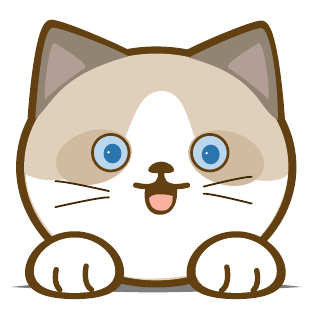}}~\textbf{\TOOLw{MCG}}: This variant investigates the impact of the \mcg. Only the \mcg view is provided on top of \TOOLbase, giving the agent an overview of coupled module regions to orient and narrow its search space.
\item \raisebox{-0.20em}{\includegraphics[height=1em]{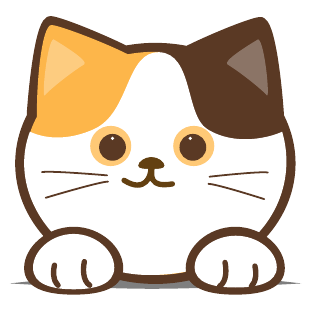}}~\textbf{\TOOLw{FCG}}: This variant explores the impact of the \fcg. Only the \fcg view is provided on top of \TOOLbase, exposing caller-callee relationships around a function, assisting the agent for interprocedural reasoning.
\item \raisebox{-0.20em}{\includegraphics[height=1em]{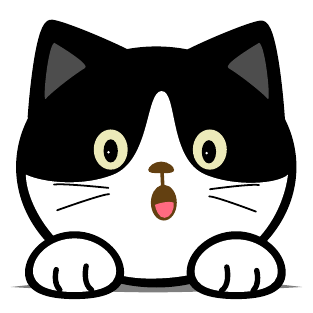}}~\textbf{\TOOLw{CHG}}: This variant explores the impact of the \chg. Only the \chg view is provided on top of \TOOLbase, displaying the inheritance and overriding relations around a class of interest, helping the agent determine which implementation governs the observed behavior.
\item \raisebox{-0.20em}{\includegraphics[height=1em]{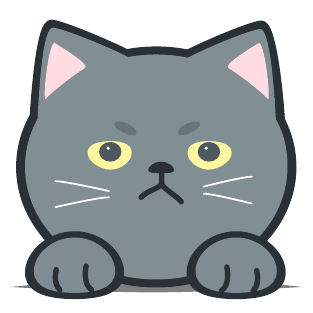}}~\textbf{\TOOLw{PDG}}: This variant explores the impact of the \pdg. Only the \pdg view is provided on top of \TOOLbase, presenting statement-level data and control dependencies within a candidate implementation, helping the agent reason about predicates, values, and state updates.
\item \raisebox{-0.20em}{\includegraphics[height=1em]{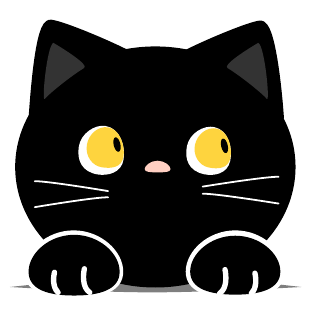}}~\textbf{\TOOLtextual}: 
This variant evaluates textual graph representations. All four graph views are serialized into text without rendered images, preserving the same structural information as \TOOLfull. Comparing it against \TOOLfull\ isolates the benefit of visual representations.

\item \raisebox{-0.20em}{\includegraphics[height=1em]{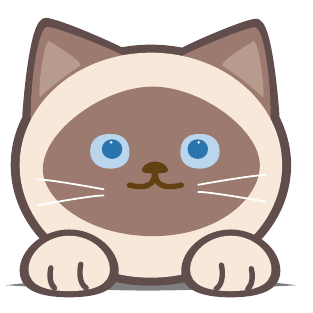}}~\textbf{\TOOLvisual}: 
This variant evaluates visual graph representations. All four graph views are rendered as images, while their textual descriptions are removed. Comparing it against \TOOLfull\ / \TOOLtextual\ isolates the benefit of textual details complementary to visual observations.

\item \raisebox{-0.20em}{\includegraphics[height=1em]{Icons/full.pdf}}~\textbf{\TOOLfull}: 
For reference, we include the complete \TOOL with four graph views (\mcg, \fcg, \chg, and \pdg) provided through both visual images and textual descriptions, enabling dual-modal structural reasoning.
\end{itemize}

\begin{table}[t]
\vspace{3pt}
\caption{
% Ablation study of graph views and visual modality in \TOOL on a subset of \swebp.
Ablation study of multi-grained graph views and dual-modal structural observations in \TOOL.
}
\label{tab:rq2}
\centering
\small
\setlength{\tabcolsep}{4pt}
\resizebox{\columnwidth}{!}{
\begin{tabular}{lccccccccc}
\toprule
\multirow{2}{*}[-0.7ex]{\textbf{Variant}} & \multicolumn{4}{c}{\textbf{Graph Views}} & \multicolumn{2}{c}{\textbf{Graph Modality}} & \multirow{2}{*}[-0.7ex]{\textbf{Resolved}} & \multirow{2}{*}[-0.7ex]{\textbf{\$Avg.\ Cost}} & \multirow{2}{*}[-0.7ex]{\textbf{\#Avg.\ Step}} \\
\cmidrule(lr){2-5}\cmidrule(lr){6-7}
 & \mcg & \fcg & \chg & \pdg & Textual & Visual & & & \\
\midrule
\raisebox{-0.20em}{\includegraphics[height=1em]{Icons/base.pdf}} \TOOLbase    & \textendash & \textendash & \textendash & \textendash & \textendash & \textendash & 69 & \$0.54 & 49.3 \\
\midrule
\raisebox{-0.20em}{\includegraphics[height=1em]{Icons/mcg.pdf}} \TOOLw{MCG} & \checkmark & \textendash  & \textendash  & \textendash  & \checkmark  & \checkmark & 72 & \$0.53 & 45.1 \\
\specialrule{0pt}{2pt}{0pt}
\raisebox{-0.20em}{\includegraphics[height=1em]{Icons/fcg.pdf}} \TOOLw{FCG} & \textendash  & \checkmark & \textendash  & \textendash  & \checkmark  & \checkmark & 74 & \$0.58 & 51.7 \\
\specialrule{0pt}{2pt}{0pt}
\raisebox{-0.20em}{\includegraphics[height=1em]{Icons/chg.pdf}} \TOOLw{CHG} & \textendash  & \textendash  & \checkmark & \textendash  & \checkmark  & \checkmark & 73 & \$0.53 & 48.3 \\
\specialrule{0pt}{2pt}{0pt}
\raisebox{-0.20em}{\includegraphics[height=1em]{Icons/pdg.pdf}} \TOOLw{PDG} & \textendash  & \textendash  & \textendash  & \checkmark & \checkmark  & \checkmark & 75 & \$0.59 & 51.6 \\
\midrule
\raisebox{-0.20em}{\includegraphics[height=1em]{Icons/text.pdf}} \TOOLtextual    & \checkmark  & \checkmark  & \checkmark  & \checkmark  & \checkmark  & \textendash & 71 & \$0.57 & 51.1 \\
\specialrule{0pt}{2pt}{0pt}
\raisebox{-0.20em}{\includegraphics[height=1em]{Icons/visual.pdf}} \TOOLvisual    & \checkmark  & \checkmark  & \checkmark  & \checkmark  & \textendash  & \checkmark  & 78 & \$0.50 & 47.2 \\
\midrule
\raisebox{-0.20em}{\includegraphics[height=1em]{Icons/full.pdf}} 
\TOOLfull    & \checkmark  & \checkmark  & \checkmark  & \checkmark  & \checkmark  & \checkmark  &  81 &  \$0.49 & 45.3 \\
\bottomrule
\vspace{-15pt}
\end{tabular}}
\end{table}

\subsubsection{Benefit of Dual-Modal Structural Reasoning}
We first compare \TOOLfull\ with \TOOLbase\ to evaluate the overall benefit of introducing dual-modal repository structure into issue resolution. As shown in Table~\ref{tab:rq2}, \TOOLfull\ resolves 81 issues, compared with 69 for \TOOLbase, yielding 12 additional fixes and a 17.4\% relative improvement. Meanwhile, the average exploration steps decrease from 49.3 to 45.3 and the average cost decreases from \$0.54 to \$0.49.
The two variants differ fundamentally in how repository relationships are exposed to the agent. \TOOLbase\ relies on conventional textual exploration tools, such as file inspection and keyword search, which reveal repository information incrementally through fragmented observations. 
% \jy{this sentence can be deleted}
% \kai{will remove it}
% As a result, the agent must repeatedly reconstruct repository structure from individual files, symbols, and search results when reasoning about cross-file dependencies. In contrast, \TOOLfull\ externalizes repository relationships as queryable graph views and exposes them through both visual and textual modalities. This allows the agent to directly access structural information rather than repeatedly inferring it from scattered textual evidence.

\begin{figure}[t]
    \centering
    {\includegraphics[width=1.0\columnwidth, trim=165 53 165 62, clip]{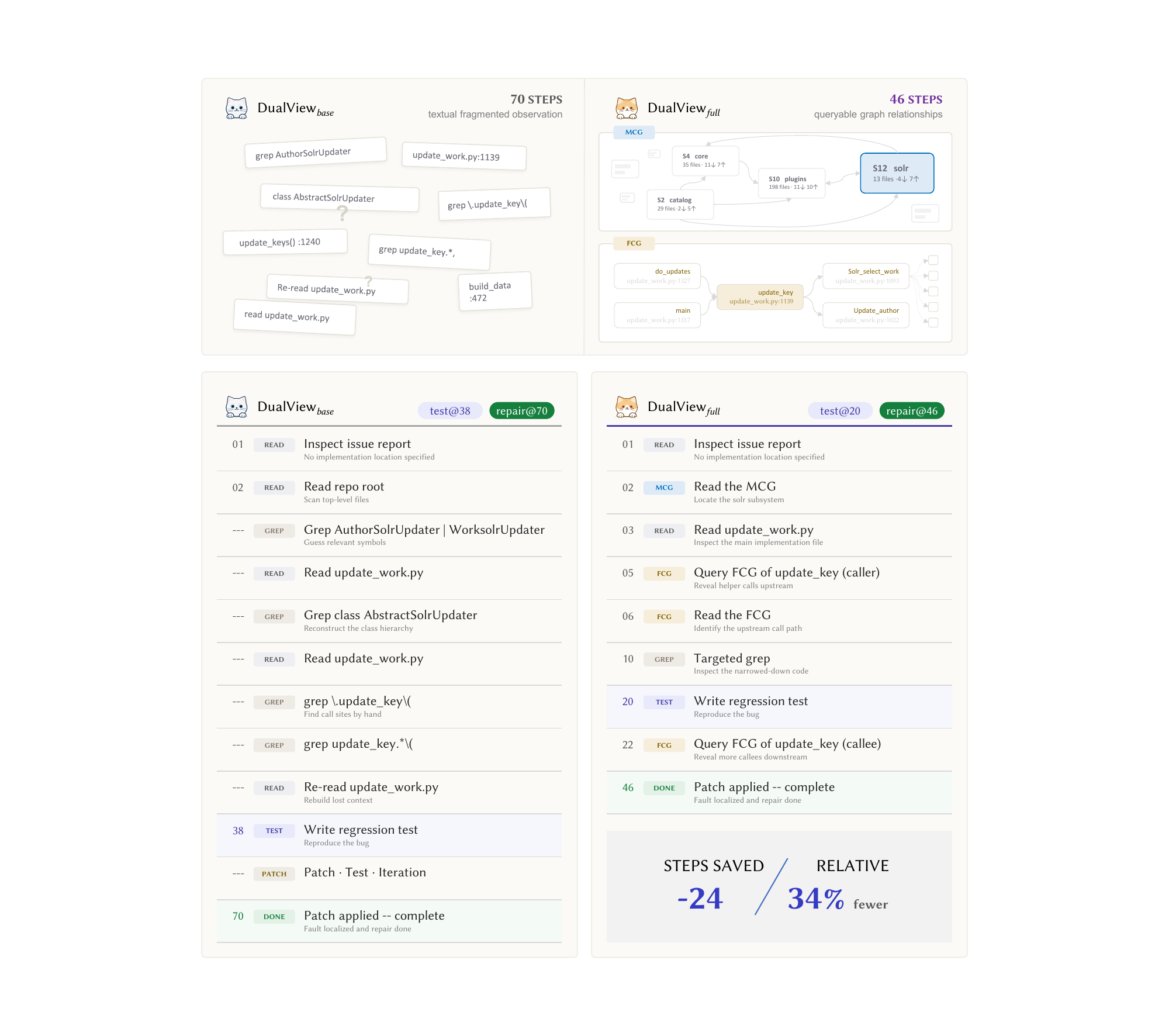}}
    \caption{The case study of \textit{openlibrary-b4f7c1}.}
    \label{fig:rq2_1_case}
    \vspace{-15pt}
\end{figure}

The case in Figure~\ref{fig:rq2_1_case} illustrates this difference. By leveraging MCG and FCG, \TOOLfull\ rapidly identifies the relevant subsystem and dependency chain, constructing the regression test after only 20 tool calls and completing the repair after 46 calls. In contrast, \TOOLbase\ requires repeated file inspection and search operations to reconstruct the same structural context, resulting in 38 calls before test construction and 70 calls before repair completion. This case shows that dual-modal structural observations serve as navigation scaffolds, reducing unnecessary exploration and mitigating Limitation~\textcolor{box4color}{\textbf{\ding{182}}}.

\subsubsection{Contribution of Visual and Textual Modalities}

We next investigate how different graph-consumption modalities contribute to issue resolution. To this end, we compare \TOOLfull\ with \TOOLvisual\ and \TOOLtextual. All three variants expose the same four graph views and therefore contain identical structural information. They differ only in how this information is presented to the agent.

As shown in Table~\ref{tab:rq2}, \TOOLvisual\ substantially outperforms \TOOLtextual, resolving 78 issues compared with 71. It also requires fewer exploration steps (47.2 vs. 51.1) and achieves a lower average cost (\$0.50 vs. \$0.57). These results indicate that visual graph renderings provide a more effective interface for repository exploration than textual graph serializations. By preserving graph topology directly, visual representations allow the model to identify dependency paths, branching structures, and highly connected regions without reconstructing them from lengthy textual descriptions. Consequently, the agent can navigate the repository more efficiently and reason about long-range dependencies with fewer exploration steps, mitigating Limitation~\textcolor{box4color}{\textbf{\ding{183}}}.

\begin{figure}[t]
    \centering
    {\includegraphics[width=1\linewidth]{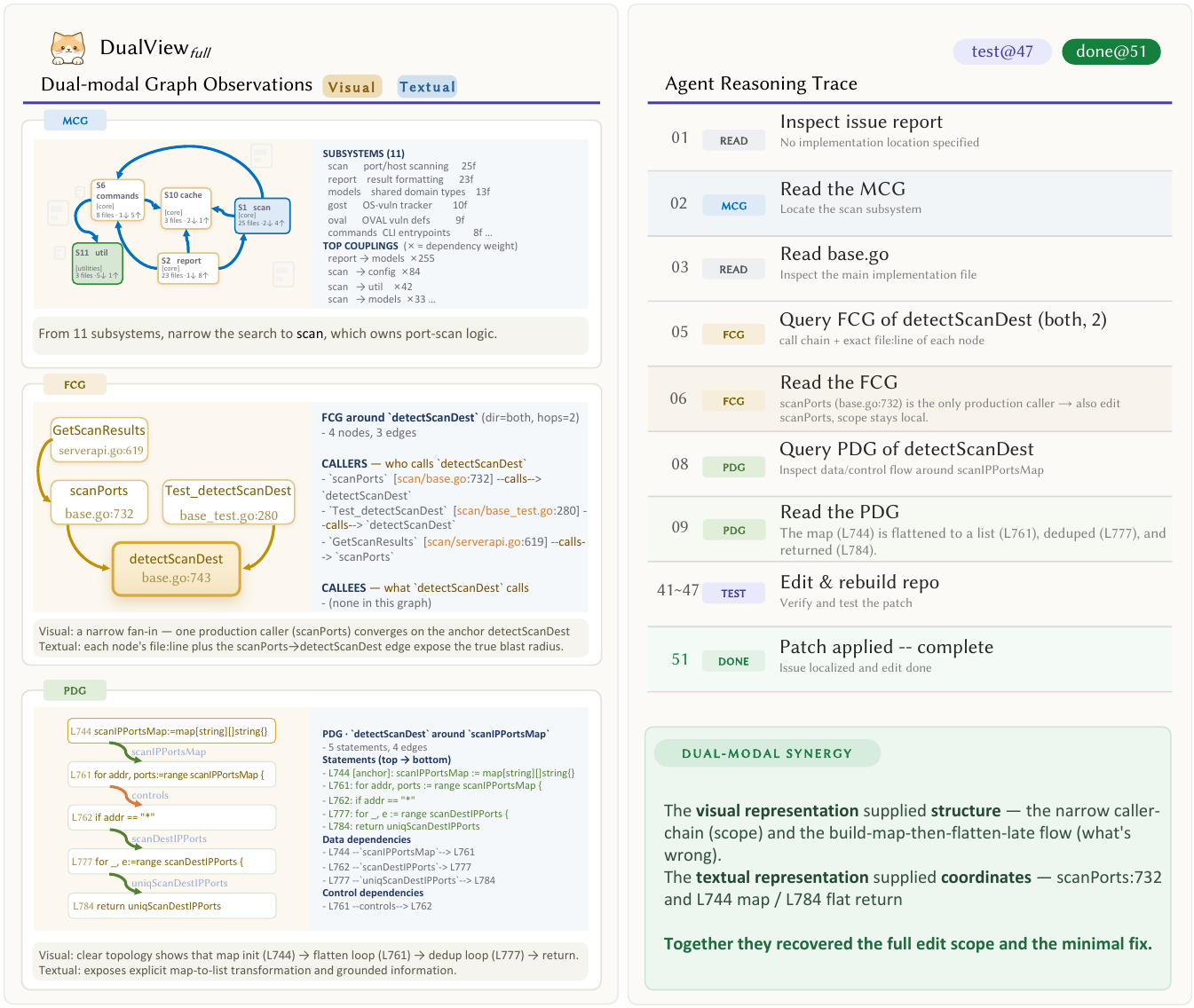}}
    \caption{The case study of \textit{vuls-edb324}.}
    \label{fig:rq2_2_case}
    \vspace{-8pt}
\end{figure}

However, visual representations alone are insufficient. Although \TOOLvisual\ performs strongly, \TOOLfull\ resolves 3 additional issues while further reducing cost and steps. 
% As shown in Figure 5, the agent resolves a refactoring issue in future-architect/vuls, where detectScanDest must return ports grouped by IP (a map[string][]string) instead of a flat ip:port slice. Although the visual and textual views encode the same graph, they help the agent in different ways. The visual graphs convey the overall shape of the code at a glance: the FCG shows that only one caller, scanPorts, uses the function, so the change stays local, while the PDG shows that the function builds the grouped map but flattens it before returning. Recovering the same understanding from a textual edge list would require rebuilding the graph mentally first. The textual views instead supply the exact code locations to edit—scanPorts at base.go:732, the map built at L744, and the flat value returned at L784—which appear in the image but are too small to read precisely, even though an edit must point to an exact line. The two modalities are thus complementary not because they carry different information, but because they present it in different forms: the visual graph reveals structure quickly and narrows the search, while the textual graph supplies the precise locations needed to act. Having both lets the agent recover the full edit scope and the minimal change more reliably than either view alone.
\jy{% This indicates that the cooperation of dual modalities can maximize the effectiveness of \TOOL. 
As shown in Figure~\ref{fig:rq2_2_case}, \TOOL guides the agent through a top-down exploration using dual-modal structural observations. The reasoning trace highlights the distinct role of each view: the agent invokes the MCG to isolate the subsystem, the FCG to confirm the local blast radius, and the PDG to uncover the underlying data and branching logic. Throughout this process, visual graphs convey structural topology at a glance, allowing the agent to reason without requiring mental reconstruction of scattered textual edges. Conversely, the textual views supply the grounded coordinates strictly required to synthesize precise edits. 
Together, they enable the agent to reliably recover the full edit scope and execute a minimal fix.}
% \kai{please add description for this figure, and try to refine this figure since it is not beautiful enough and text is too small}
% This observation suggests textual graph representations provide complementary information that is difficult to capture from images alone. 
% In particular, textual views explicitly expose node names, edge semantics, file paths, confidence levels, and dependency attributes that may be visually compressed or less salient in graph renderings. 
\kai{The two modalities are thus complementary not because they carry different information, but because they present it in different forms: the visual graph reveals structure quickly and narrows the search, while the textual graph supplies the precise locations needed to act.}
% Having both lets the agent recover the full edit scope and the minimal change more reliably than either view alone.}
During our analysis, we observed 
% \kai{several (need a detailed NUM)} 
12
trajectories where the agent used visual graphs to identify the relevant dependency region and subsequently relied on textual graph descriptions to inspect specific nodes and relations before making editing decisions.

\begin{figure}[t]
    \centering
    % \vspace{-13pt}
    \includegraphics[width=\linewidth]{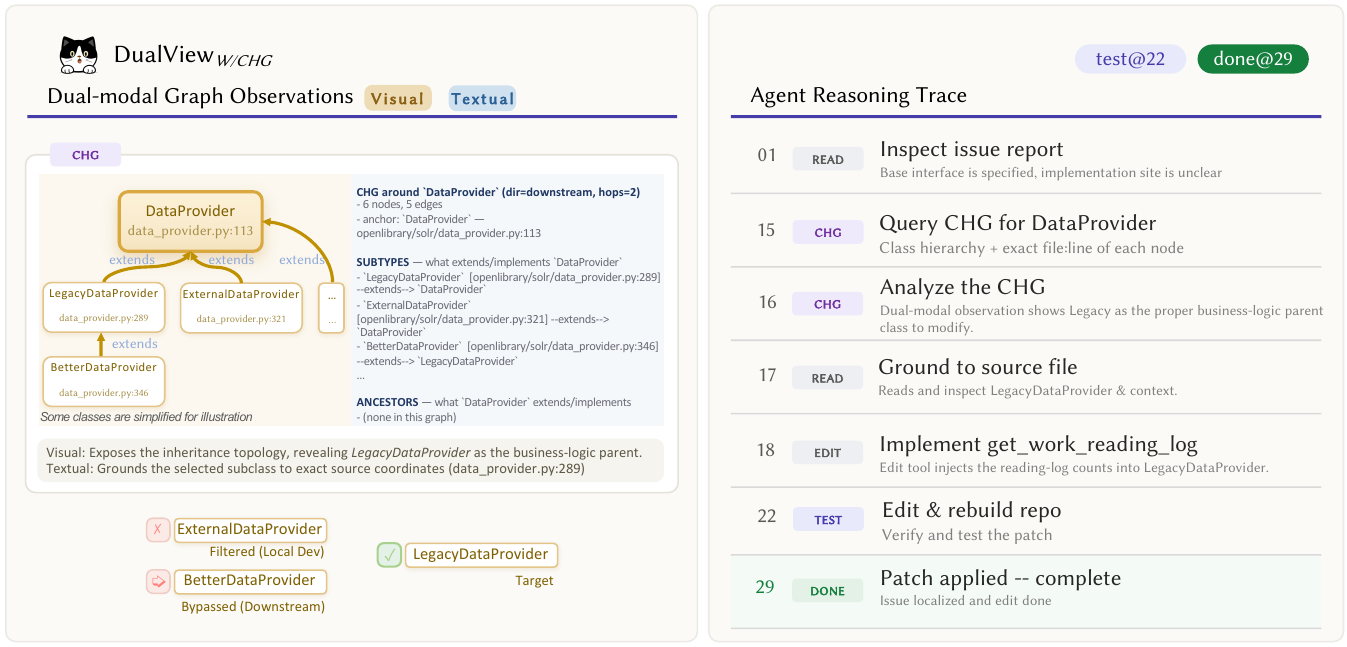}
    \caption{The CHG case study of \textit{openlibrary-7bf323}.}
    \label{fig:rq2_chg_case}
    \vspace{-15pt}
\end{figure}

\subsubsection{Contribution of Individual Graph Views}
We compare \TOOLbase\ with each single-view variant, including \TOOLw{MCG, FCG, CHG, PDG}. 
% As shown in Table~\ref{tab:rq2}, the \TOOLbase\ resolves 69 instances without access to any graph view. 
When the individual view \mcg, \fcg, \chg and \pdg is enabled, each variant consistently improves the resolution performance, achieving 72, 74, 73 and 75 instances respectively. These results show that each graph view provides useful structural evidence compared to \TOOLbase.
While Figure~\ref{fig:rq2_2_case} illustrates the contributions of \mcg, \fcg, and \pdg, the unique usage of \chg is shown in  Figure~\ref{fig:rq2_chg_case}. Consider an OpenLibrary issue requiring a new \texttt{DataProvider.get\_work\_reading\_log} method without specifying the target subclass. Querying the CHG exposed the inheritance topology, enabling the agent to strategically prune the search space. By evaluating sibling nodes simultaneously, it filters out \texttt{ExternalDataProvider} as a local-dev stub and bypasses the downstream \texttt{BetterDataProvider}. The agent thus isolated \texttt{LegacyDataProvider} as the authoritative business-logic parent, grounding its edit to exact source coordinates. The resulting patch naturally propagated via inheritance, proving CHG resolves polymorphic ambiguity and guides ``where-to-implement'' decisions without deep-diving into irrelevant implementations.
\subsubsection{Complementarity of Multiple Graph Views}
We compare \TOOLfull\ with the single-view variants \TOOLw{MCG, FCG, CHG, PDG}. \TOOLfull\ (81 resolved) outperforms these single-view variants (72-75 resolved). The improvement over individual views indicates that no single representation is sufficient for all issue-resolution scenarios. Instead, graph views at different granularities provide complementary evidence: \mcg supports repository-level orientation, \fcg captures interprocedural propagation, \chg exposes dispatch relations, and \pdg reveals intra-function dependencies. 
% In a representative issue shown in Figure~\ref{rq2_full_case}, \TBD{describe how the agent first uses a coarse-grained graph to identify the relevant subsystem and subsequently invokes one or more fine-grained views to locate and validate the affected implementation}. This trajectory illustrates how multiple graph views jointly support long-horizon exploration by preserving relevant structural context across successive reasoning steps. 
% \jy{We can say that the case in Figure~\ref{fig:rq2_1_case} has already reflected the complementarity of multiple graphs (MCG and FCG).}
Figure~\ref{fig:rq2_1_case} illustrates such collaboration between graph views: the agent uses MCG to establish the relevant repository context and then invokes FCG to reason about call dependencies around the target function. This multi-view trajectory shows why \TOOLfull\ can outperform single-view variants: long-horizon issue resolution often requires the agent to switch between structural views as its reasoning focus changes.

% \end{enumerate}

\begin{figure*}[h]
    \vspace{-6pt}
    \centering
    \includegraphics[width=\linewidth]{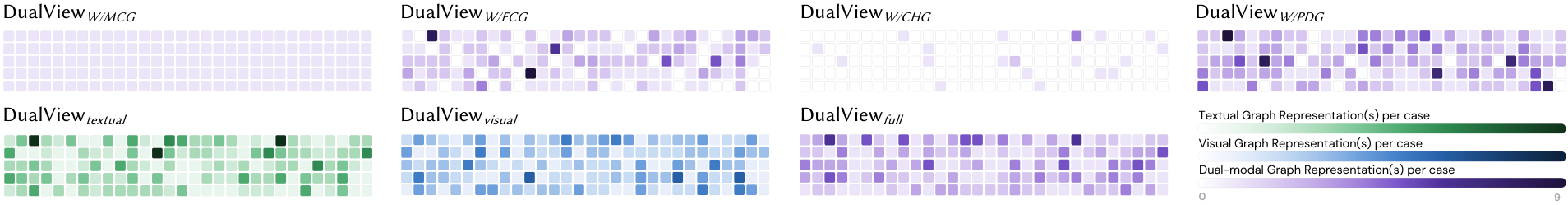}
    \caption{Heatmap of different variants of \TOOL invoking corresponding graph representation(s) frequency. \chen{Too many different cats, cannot recognize which one is what approach, suggest to add the full name on the top of each subfigure.}}
    \label{fig:rq2_heatmap}
    \vspace{-6pt}
\end{figure*}

Figure~\ref{fig:rq2_heatmap} further characterizes how \TOOL accesses graph views during issue resolution. Each cell represents one issue in the evaluated subset, and its intensity indicates the number of accesses to the corresponding graph representation. The \mcg overview is typically consumed once at the beginning of a trajectory to provide repository-level orientation. In contrast, \fcg, \chg, and \pdg are requested more selectively, and some issues can be resolved without invoking additional fine-grained views. For structurally complex issues, the agent may revisit these views multiple times as the exploration proceeds. These heterogeneous usage patterns show that \TOOL does not indiscriminately inject all graph representations into every trajectory. Instead, it exposes structural evidence on demand, allowing the agent to retrieve relevant repository relationships during long-horizon exploration.

% Figure~\ref{fig:rq2_heatmap} further shows that graph views are invoked selectively rather than uniformly across issues. MCG is typically accessed once to establish repository-level context, whereas FCG, CHG, and PDG are queried on demand according to the evolving repair trajectory. The heterogeneous usage patterns indicate that \TOOL functions as a queryable structural scaffold rather than a static context augmentation mechanism, enabling agents to retrieve relevant structural evidence only when needed.

% Overall, the ablation results show that each graph view improves issue-resolution performance, while combining them produces the strongest results. The comparison between \TOOLtext and \TOOLfull\ further demonstrates that visual representations provide benefits beyond the availability of structural information alone. The heatmap proves agent can invoke corresponding the graph at different level based on the need of the issue being resolved. Together, these findings support our central design rationale: explicitly visualizing repository relationships helps agents reconstruct complex dependency structures and provides reusable structural context for exploring long-horizon codebases more efficiently.

% \scriptsize
\subsubsection{Summary}
% \textbf{:}
The ablation study provides empirical evidence for both design dimensions of \TOOL: 
\textbf{1) Dual-modal observations.}
Compared with text-only repository exploration, visualized repository structures substantially improve issue-resolution performance. Moreover, visual graph representations outperform textual graph serializations, while the combination of visual and textual graph modalities achieves the strongest results, demonstrating their complementary strengths for structural reasoning.
\textbf{2) Multi-grained views.}
Each graph view contributes useful structural evidence beyond the base agent, and combining multiple views consistently outperforms any individual view. This result confirms that repository exploration benefits from structural information at different granularities, ranging from repository-level organization to fine-grained dependency analysis.

% \jy{Overall, these findings confirm our core premise: navigating effective repository reasoning requires both complementary multi-grained structural views and dual-modal consumption.}
% Overall, these findings support the central premise of \TOOL: effective repository reasoning requires both complementary graph views and complementary modalities. By combining multi-grained structural representations with dual-modal graph consumption, \TOOL enables more effective navigation and reasoning over long-horizon repository dependencies.

\subsection{RQ3: Generalizability Study}
\label{sec:eval:rq3}

\begin{table}[]
\vspace{4pt}
\caption{Resolved instances on SWE-bench Verified.}
\label{tab:result_RQ3_gen_study}
\resizebox{1.0\columnwidth}{!}{
\begin{tabular}{llccc}
\toprule
\multicolumn{1}{c}{\textbf{Agent System}}          & \multicolumn{1}{c}{\textbf{Base Model}} & \textbf{Resolved} & \textbf{\%Resolved} & \textbf{\$Avg. Cost} \\
\midrule
mini-SWE-agent & Claude 4.5 Opus (high reasoning) & 384 & 76.8\% & \$0.75 \\
mini-SWE-agent & Gemini 3 Flash (high reasoning)  & 379 & 75.8\% & \$0.36 \\
mini-SWE-agent & MiniMax M2.5 (high reasoning)    & 379 & 75.8\% & \$0.07 \\
mini-SWE-agent & Claude 4.6 Opus                  & 378 & 75.6\% & \$0.55 \\
mini-SWE-agent & GPT-5-2 Codex                    & 364 & 72.8\% & \$0.45 \\ 
\rowcolor[HTML]{DFDCEF}
\raisebox{-0.20em}{\includegraphics[height=1em]{Icons/full.pdf}} \TOOLat{\miniswea} & Gemini 3 Flash (high reasoning)         & 399                 & \textbf{79.8\%}              & \$0.18                    \\ \bottomrule
\end{tabular}
}
\vspace{-6pt}
\end{table}

% \subsection{Generalizability on SWE-bench Verified}
\label{sec:generalizability}

To evaluate whether the effectiveness of \TOOL generalizes beyond the benchmark and model settings considered in our main experiments, we further evaluate it on SWE-bench Verified~\cite{SWEBenchVerfied}.
% , a widely used benchmark for repository-level issue resolution. 
Since many competitive results on its leaderboard are obtained with mini-SWE-agent~\cite{mini_SWE_agent}, we integrate \TOOL into the same scaffold to conduct a fair comparison. We select Gemini 3 Flash as the base model because it achieves a favorable balance between effectiveness and cost among the reported mini-SWE-agent configurations while supporting the multimodal reasoning required by \TOOL. 
% Notably, although MiniMax M2.5 is  competitive and inexpensive as well, it does not support visual inputs. 
As shown in Table~\ref{tab:result_RQ3_gen_study}, \TOOL resolves 399 out of 500 instances, achieving a resolution rate of 79.8\%. Compared with the original mini-SWE-agent using the same model, \TOOL resolves 20 additional instances and improves the resolution rate from 75.8\% to 79.8\%, while reducing the average cost from \$0.36 to \$0.18 per instance. 
These results show that the effectiveness of \TOOL generalizes to an additional benchmark and a lightweight multimodal model, while improving both resolution accuracy and cost efficiency.

% \jy{The major overhead difference here is maybe because we enabled prompt caching}
% Moreover, \TOOL outperforms all listed mini-SWE-agent configurations, including Claude 4.5 Opus with high reasoning, which resolves 384 instances at an average cost of \$0.75. 
\section{Threats to Validity}
\label{sec:threats}

\chen{this section can be removed if more space is needed.}
% \textbf{Data leakage.}
% The repositories in \swebp\ and \swebv\ are publicly available and may overlap with the pretraining data of modern LLMs. To mitigate this threat, all comparisons use identical base models across \TOOL and the corresponding baselines. The consistent gains achieved by \TOOL under the same models suggest that the improvements stem from visual structural reasoning rather than solely from memorized repository knowledge.

\noindent
\textbf{Experimental variance.}
LLM-based agents are inherently nondeterministic, while each benchmark instance in our evaluation was executed once due to the high computational cost of repository-level issue resolution. Although this follows common practice in recent SWE-bench evaluations, future work could further reduce potential variance by averaging results over multiple independent runs.

\noindent
\textbf{Visual rendering.}
The effectiveness of visual reasoning may depend on how repository structures are rendered. To reduce this threat, all graph views are generated using deterministic Graphviz layouts with fixed rendering parameters across all experiments. Although these settings may not be optimal, \TOOL consistently improves repair performance without renderer-specific tuning, suggesting that our conclusions are robust to the chosen visualization configuration.

\section{Related Work}
\label{sec:related}

% Agentic Program Repair for Issue Resolution
% Program Repair研究在近年来进入来agent时代，早期的研究人员通过设计agent系统来来借助测试用例的驱动来执行调试和修复。然而，这些工作要么是建立在完美缺陷定位的假设之上或者依赖测试用例的反馈驱动修复。最近，SWE-Bench家族的benchmarks的提出将APR研究转移到解决真实世界的github issues上去。这要求agent系统需要从头开始探索代码库通过实施一系列行为来实现理解issue根因，实施缺陷定位，收集修复成分，生成测试用例，合成修复行为等。然而，目前的工作在探索代码库方面仍然仅仅只是通过碎片化的文本观察来重建对于复杂依赖关系的理解从而实施代码库探索和导航。例如,mini-swe-agent像许多CLI agent那样通过执行各种命令行的命令来探索代码库或执行各种行为。
% 不同于这些工作，DualView设计了额外的视觉化接口来对程序内部复杂的代码依赖关系进行建模，从而避免agent系统从碎片化的代码片段中重建依赖分析的....，并允许agent在现有基础能力上进一步增强视觉化接口来增强代码库探索能力。(这里有点想解决limitation 1的意思)
% 最近一些工作也开始关注如何提取代码库内部的关系实施建模从而帮助agent探索代码库。例如，Prometheus通过 represents the repository as a knowledge graph To facilitate semantic understanding and context retrieval.
% 不同于这些工作利于图关系来实施context检索的碎片化线索收集，DualView强调将文本化的图关系建模转向视觉化的图关系呈现从而更好地帮助agent解决视觉化线索来实施空间导航，这能避免agent需要从文本化图关系描述中分析并推理依赖关系的繁琐，而是直接将文本化的复杂线索抽象到更高层的视觉化直接呈现来让agent直接推理。(这里有点想解决limitation 2的意思)

% Kai: I provide a short version
% ---------------------------------

\chen{this section can be removed if more space is needed.}
% \kai{Just meet papge limit}

\noindent
\textbf{Agentic Program Repair for Issue Resolution.}
% Recent advances in LLM-based agents have shifted automated program repair from patch generation toward repository-level issue resolution. 
Driven by the SWE-bench family of benchmarks~\cite{SWEBench,SWEBench_Pro,SWEBench_M}, a variety of issue-resolution agents have been proposed, including SWE-agent~\cite{SWE_Agent}, AutoCodeRover~\cite{zhang2024autocoderover}, OpenHands~\cite{OpenHands}, etc. These systems improve repair performance through tool use, retrieval mechanisms, planning strategies, and workflow design. However, repository exploration remains primarily based on textual observations, such as source files, search results, and command outputs. In contrast, \TOOL augments existing issue-resolution agents with visual structural representations that make repository relationships directly observable.

\noindent
\textbf{Repository Structure Modeling for Agent Reasoning.}
Recent work has explored modeling repository structure to support agent reasoning. Representative approaches construct repository graphs, code graphs, or knowledge graphs to capture dependencies among files, functions, classes, and symbols, enabling more effective retrieval and context acquisition~\cite{CodeGraph,RepoGraph,CoSIL,ARISE,Prometheus,KGCompass,SGagent}. While these approaches provide valuable structural information, the resulting relationships are typically consumed through retrieval results, graph serialization, or symbolic queries. In contrast, \TOOL treats repository structure as a first-class reasoning artifact by externalizing repository relationships into visual graph representations and integrating them directly into the agentic reasoning loop.

\noindent
\textbf{Visual Reasoning for Code Comprehension.}
Recent MLLMs have demonstrated 
% strong 
visual reasoning capabilities, motivating researchers to explore visual representations for SE tasks. Prior work has leveraged 
% screenshots and 
UI artifacts for 
% multmodal
visual
issue resolution~\cite{GUIRepair,SVRepair}, while 
% more 
recent studies investigate representing source code itself as images to improve code understanding and long-context processing~\cite{LongCodeZip,CodeOCR,LongCodeOCR}. Our work differs from these efforts in both representation and objective. Rather than visualizing raw code, \TOOL visualizes repository structure through SE graphs that expose module dependencies, call relationships, inheritance hierarchies, and program dependencies. This enables agents to reason directly over repository topology and long-range dependency chains during issue resolution.

\section{Conclusion}
\label{sec:conclusion}

We presented \TOOL, a dual-modal structural reasoning framework that exposes repository structure through complementary visual and textual graph representations. Our results show that repository graphs serve more effectively as reasoning artifacts than retrieval artifacts, highlighting the potential of multimodal repository reasoning for agent systems.

\bibliographystyle{IEEEtran}
\bibliography{main}

\end{document}